\documentclass[aps, reprint, prl, twocolumn, longbibliography]{revtex4-1}
\usepackage{graphicx}
\usepackage{amsmath}
\usepackage{bm}
\usepackage{braket}
\usepackage{psfrag} 
\usepackage{latexsym} 
\usepackage{amstext}
\usepackage{amsxtra} 
\usepackage{float}
\usepackage{upgreek} 
\usepackage{times} 
\usepackage[bookmarks=false]{hyperref}
\usepackage{pgffor}
\usepackage{pdfpages}        
\usepackage{lineno}
\hypersetup{
    colorlinks=true,
    linkcolor=blue,
    urlcolor=blue,
    citecolor=blue,
    filecolor=blue
    }
\allowdisplaybreaks

\makeatletter
\AtBeginDocument{\let\LS@rot\@undefined}
\makeatother

\AtBeginDocument{%
  \setlength{\linenumbersep}{0.1cm}%
}

\begin{document}

\newcommand{\kB}{k_{\rm B}}
\newcommand{\cs}{$\clubsuit \; $}
\newcommand{\downstate}{\left\vert\downarrow\right\rangle}  
\newcommand{\upstate}{\left\vert\uparrow\right\rangle}
\newcommand{\overbar}[1]{\mkern 1.5mu\overline{\mkern-1.5mu#1\mkern-1.5mu}\mkern 1.5mu}
\newcommand{\expect}[1]{\langle#1\rangle}
\newcommand{\lb}{\ell_B}
\newcommand{\vh}{v_\textrm{d}}
\newcommand{\vhVec}{\vec{v}_\textrm{d}}
\newcommand{\affcua}{MIT-Harvard Center for Ultracold Atoms, Research Laboratory of Electronics, and Department of Physics, Massachusetts Institute of Technology, Cambridge, Massachusetts 02139, USA}
\newcommand{\Vs}{V_\textrm{s}}
\newcommand{\sLLL}{\sigma_\textrm{LLL}}
\newcommand{\nll}{N_\textrm{LL}}
\newcommand{\ks}{k_\textrm{max}}
\newcommand{\rtf}{R_\textrm{TF}}
\newcommand{\omc}{\omega_\textrm{c}}

\title{
Crystallization of Bosonic Quantum Hall States
}

\author{Biswaroop Mukherjee, Airlia Shaffer, Parth B. Patel, Zhenjie Yan, Cedric C. Wilson, Valentin Cr\'epel, Richard J. Fletcher, Martin Zwierlein}
\affiliation{\affcua}

\date{\today}



\maketitle
{ \textbf {
The dominance of interactions over kinetic energy lies at the heart of strongly correlated quantum matter,
from fractional quantum Hall liquids~\cite{Stormer:1999}, to atoms in optical lattices~\cite{Bloch:2008} and twisted bilayer graphene~\cite{Cao2018}.
Crystalline phases often compete with correlated quantum liquids, and transitions between them occur when the energy cost of forming a density wave approaches zero. A prime example occurs for electrons in high magnetic fields, where the instability of quantum Hall liquids towards a Wigner crystal~\cite{Wigner:1934,Yoshioka1979,Lam:1984, Girvin:1985,Girvin:1986,Goldman:1990, Jiang:1990,Jang2017Wigner} is heralded by a roton-like softening of density modulations at the magnetic length~\cite{Haldane1985,Girvin:1985,Girvin:1986,Pinczuk1993,Kukushkin2009}.
Remarkably, interacting bosons in a gauge field are also expected to form analogous liquid and crystalline states ~\cite{Cooper:2001,Ho:2001,Oktel:2004,Sinha:2005,aftalion:2009,Chen:2012,Lu2012,Senthil:2013,Vishwanath:2013,dalibard:2011,goldman:2014,galitski:2019}. However, combining interactions with strong synthetic magnetic fields has been a challenge for experiments on bosonic quantum gases~\cite{dalibard:2011,goldman:2014, galitski:2019}. 
Here, we study the purely interaction-driven dynamics of a Landau gauge Bose-Einstein condensate~\cite{Fletcher:2019} in and near the lowest Landau level (LLL).
We observe a spontaneous crystallization driven by condensation of magneto-rotons~\cite{Haldane1985,Girvin:1985,Girvin:1986}, excitations visible as density modulations at the magnetic length.
Increasing the cloud density smoothly connects this behaviour to a quantum version of the  Kelvin-Helmholtz hydrodynamic instability, driven by the sheared internal flow profile of the rapidly rotating condensate.
At long times the condensate self-organizes into a persistent array of droplets, separated by vortex streets, which are stabilized by a balance of interactions and effective magnetic forces.
}
}

When electrons are placed in a magnetic field, their kinetic energy is quenched. The single particle states form discrete, highly degenerate Landau levels, and correspond to wavepackets localized to the magnetic length $\lb$. 
In the presence of interactions between electrons, owing to the absence of kinetic energy, one naturally expects the formation of a Wigner crystal of periodicity ${\sim \lb}$~\cite{Wigner:1934, Yoshioka1979,Tsui1982, Lam:1984,Goldman:1990,Jiang:1990,Jang2017Wigner}.
Famously however, the interplay of the macroscopic degeneracy and interactions instead typically favours the strongly correlated fractional quantum Hall liquids, which host fractional charges, anyonic exchange statistics, and topologically protected transport properties~\cite{Stormer:1999}. The tendency to crystallize is still apparent in a pronounced minimum in the collective excitation spectrum at wavevectors ${k~{\sim}~ 1/\lb}$~\cite{Haldane1985,Girvin:1985,Girvin:1986,Pinczuk1993,Kukushkin2009}. In analogy with the roton minimum in $^4$He, also considered a precursor of solidification~\cite{Nozieres2004}, these excitations are called magneto-rotons~\cite{Girvin:1985,Girvin:1986,Pinczuk1993,Kukushkin2009}.

The fate of interacting bosons in the presence of a gauge field is of fundamental importance in the classification of topological states of matter~\cite{Chen:2012,Lu2012}.
 Quantum Hall states~\cite{Cooper:2001, Ho:2001,Senthil:2013,Vishwanath:2013}, exotic vortex lattices~\cite{Oktel:2004} and vortex-free states under extreme fields~\cite{aftalion:2009} were predicted. Quantum phase transitions between such states were found to be signaled by the softening of a roton-like collective mode~\cite{Sinha:2005}.

Bosonic quantum gases in artificial magnetic fields~\cite{dalibard:2011,goldman:2014, galitski:2019} have been generated via spin-orbit coupling~\cite{Lin:2009c,Galitski:2013,Chalopin2020}, phase imprinting in lattices~\cite{struck:2012,aidelsburger:2013, miyake:2013,jotzu:2014,aidelsburger:2014,celi:2014,stuhl:2015, mancini:2015}, and by rotation of the trapped gas~\cite{Schweikhard:2004a,Bretin:2004,Cooper:2008}.
The latter approach uses the analogy between the Lorentz force on a charged particle in a magnetic field, and the Coriolis force on a massive particle in a frame rotating at frequency $\Omega$, giving ${\omc=2\Omega}$ and ${\lb=\sqrt{\hbar/(m\omc)}}$ as the rotational analog of the cyclotron frequency and the magnetic length, respectively.

Signatures of physics near the lowest Landau level have been observed in rotating Bose gases~\cite{Schweikhard:2004a,Bretin:2004}.
In recent work at MIT, condensates have been prepared directly in the lowest Landau gauge wavefunction using geometric squeezing~\cite{Fletcher:2019}.
In this mean-field quantum Hall regime~\cite{Ho:2001}, all bosons occupy a single wavefunction, whose subsequent dynamics subject to a gauge field can be studied, offering a microscopic insight into the individual building blocks of quantum Hall systems. 
An advantage of rotation is that the interactions between atoms are decoupled from the induced gauge potential, in contrast to other methods for which the effective magnetic field appears within a dressed-atom picture, leading to additional unwanted interaction terms~\cite{Bukov:2015}.

\begin{figure}[h] 
   \centering
   \includegraphics[width=1.0\columnwidth]{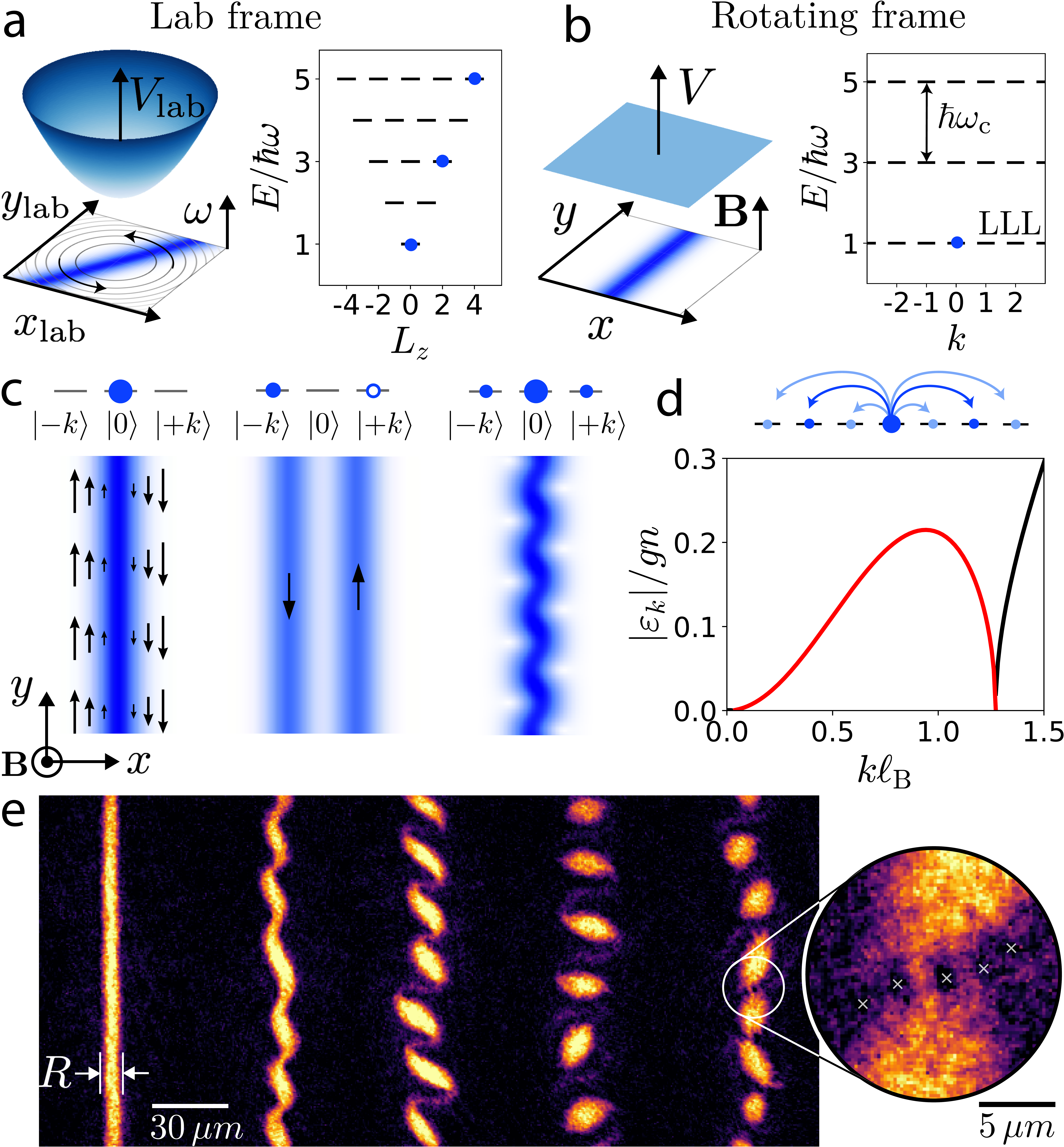} 
   \caption{
   \textbf{Spontaneous crystallization of an interacting Bose-Einstein condensate in an artificial magnetic field.} 
   \textbf{(a)}~In the laboratory frame, the condensate freely rotates in a circularly symmetric harmonic trap at the trapping frequency $\omega$. Occupied states in the energy spectrum are sketched ($L_z$: angular momentum).
   \textbf{(b)}~In the rotating frame, the condensate experiences an effective magnetic field $B$ but no scalar potential. The energy spectrum is flattened into Landau levels ($k$: momentum along $y$). Only the ${k=0}$ Landau gauge wavefunction is occupied.
   \textbf{(c)}~
   The irrotationality of the condensate in the laboratory frame imposes a sheared velocity profile in the rotating frame which is dynamically unstable towards a periodic density modulation. Motion with momentum $\hbar k$ along the $y$-direction is tied to sideways displacement of the wavefunction along $x$. The reduced overlap of ${|k|>0}$ states with the ${k=0}$ condensate lowers the interaction energy cost of collective excitations, leading to spontaneous population of ${\pm k}$ pairs whose interference with the condensate results in a density modulation. 
   \textbf{(d)}~This dynamical instability is reflected in a (Goldstone) collective excitation branch which is imaginary across a range of wavevectors, shown by a red line. The spectrum shown is calculated for a condensate in the LLL, for which the interaction energy $gn$ provides the only relevant energy scale and the magnetic length ${\ell_B=1.6\,\mu\mathrm{m}}$ sets the length scale. Here $g$ is the mean-field coupling constant, and ${n=n_{2\mathrm{D}}(0)}$ is the peak 2D density.
   \textbf{(e)}~Absorption images of the evolution of the condensate density in the rotating frame, displaying a snake-like instability and the formation of droplet arrays. Here, the cloud width $R=2.34\,\lb$, and the frames are taken in intervals of $2.5$ cyclotron periods ($2\pi/\omega_c=5.6\,\rm ms$). The zoom-in reveals vortex streets between adjacent droplets, indicating counterflow at their interface.
   }
   \label{fig:introfig}
\end{figure}

Here, we directly observe the evolution of an interacting Bose-Einstein condensate occupying a single Landau gauge wavefunction in the LLL~\cite{methods}.
We find that the Landau gauge condensate is unstable under the influence of interactions,
exhibiting spontaneous growth of a snaking mode leading to a persistent density wave order at the magnetic length $\lb$ as illustrated in Fig.~\ref{fig:introfig}. 
At the heart of this crystallization is the coupling between the relative momentum and spatial overlap of two particles in a gauge field.
This lowers the interaction energy cost of populating higher-momentum states, and leads to the dynamical instability of the lowest (Goldstone) collective excitation branch~\cite{Sinha:2005,SI}. The ensuing proliferation of excitations at momenta near $\hbar/l_B$ can be viewed as condensation of magneto-rotons, in analogy to the Wigner crystal instability of quantum Hall systems~\cite{Wigner:1934, Yoshioka1979, Lam:1984, Girvin:1985,Girvin:1986,Goldman:1990, Jiang:1990,Jang2017Wigner}.
Condensation at non-zero momentum has been predicted in superfluid helium above a critical velocity~\cite{Iordanski1980,Pitaevskii:1984}. Roton-like excitations and instabilities in Bose-Einstein condensates were induced via cavity mediated interactions~\cite{Mottl2012,Leonard:2017}, spin-orbit coupling~\cite{Ji:2015,Li:2017}, shaken optical lattices~\cite{Ha:2015,Feng:2018}, driven interactions~\cite{Zhang2020} and dipolar interactions~\cite{Petter2019, Hertkorn2021,Schmidt2021}. These instabilities were tightly connected to evidence for supersolidity, the simultaneous existence of spatial and superfluid order~\cite{Leonard:2017,Li:2017,Bottcher:2019,Chomaz:2019,Tanzi2019supersolid,Guo:2019}.
In our case, the instability to density-wave order arises purely from the interplay of contact interactions and a gauge field. No external drive is present, nor is there any residual scalar potential in the rotating frame.
The absence of kinetic energy in the LLL directly implies that the crystallization rate is set solely by the interaction energy of the gas.

By increasing the condensate density such that many Landau levels become populated, we observe a crossover from LLL behaviour to a hydrodynamic instability driven by the sheared internal velocity profile.
Analogous phenomena are ubiquitous throughout hydrodynamics, from the diocotron instability in charged plasmas~\cite{Davidson1991} and fragmentation of electron beams~\cite{Cerfon:2016}, to the Kelvin-Helmholtz instability in atmospheric and astrophysical systems~\cite{Chandrasekhar:1961,land87fluid}. 
In the context of superfluids, for which the circulation is quantized, the Kelvin-Helmholtz instability has been detected in liquid helium~\cite{Blaauwgeers:2002}, and theoretically predicted at the boundary between counterflowing condensates~\cite{Takeuchi:2010,Baggaley:2018}.
In our quantum hydrodynamic setting, we directly observe streets of quantized vortices separating emergent droplets, revealing the quantum nature of the instability at the most microscopic level.

To analyze the instability, consider the condensate in the frame rotating at the frequency $\omega$ of the isotropic harmonic trap, where it experiences a synthetic magnetic field but no scalar potential (see Fig.~\ref{fig:introfig}a,b), and thus evolves under the Hamiltonian
\begin{equation}
\hat{H}=\int\textrm{d}^2{r}\,\hat{\Psi}^\dagger\left[\frac{\left(\hat{\textbf{p}}-q\textbf{A}\right)^2}{2m}+\frac{g}{2}\hat{\Psi}^\dagger\hat{\Psi}\right]\hat{\Psi}.
\end{equation}
Here ${\hat{\Psi}^\dagger(\textbf{r})}$ is the bosonic field operator, $\hat{\textbf{p}}$ is the canonical momentum, $q$ and $\textbf{A}$ are the charge and vector potential in the equivalent magnetic problem, and $g$ is the two-dimensional mean-field coupling constant~\cite{methods}.
Geometric squeezing prepares a translationally-invariant condensate most conveniently described within the Landau gauge ${q\textbf{A}=(0,m\omega_c x)}$~\cite{Fletcher:2019,methods} for which the Hamiltonian becomes
\begin{align}
\hat{H}=\int\textrm{d}^2r\,\hat{\Psi}^\dagger\left[\frac{\hat{p}_x^2}{2m}+\frac{1}{2}m\omc^2\left(\hat{x}-\frac{\hat{p}_y\lb^2}{\hbar}\right)^2+\frac{g}{2}\hat{\Psi}^\dagger\hat{\Psi}\right]\hat{\Psi}.
\label{eqn:LandauHam}
\end{align}
Cyclotron motion of the atoms is reflected in an effective harmonic oscillator along the $x$-direction of frequency ${\omc=2\omega}$, whose non-interacting energy states correspond to different Landau levels (see Fig.~\ref{fig:introfig}b). Each level is macroscopically degenerate since it costs no energy to translate the centers of cyclotron orbits.
Initially, the $y$-momentum of all atoms is zero, and their cyclotron motion centred at ${x=0}$ with a two-dimensional number density $n_\textrm{2D}(x)$.
The condensate has uniform phase and thus features a sheared velocity profile ${\textbf{v} = -q\textbf{A}/m = \left(0,-\omega_c x\right)}$ proportional to the vector potential (see Fig.~\ref{fig:introfig}c).
We parameterize the crossover from LLL to hydrodynamic behaviour by the ratio $\frac{gn}{\hbar\omc}$ of the condensate's mean-field energy ${\sim gn}$ to the Landau level spacing ${\hbar\omega_c}$, giving a measure for the number of occupied Landau levels~\cite{Schweikhard:2004a,Fletcher:2019}. Here ${n=n_\textrm{2D}(0)}$ is the peak density. In our experiment $\frac{gn}{\hbar\omc}$ varies from $0.6$ to $7.3$, corresponding to a central filling fraction $n\lb^2$ of $50$ and higher, meaning the condensate lies within the mean-field quantum Hall regime~\cite{Cooper:2001,Ho:2001,Cooper:2008}.

The dynamical instability illustrated in Fig.~\ref{fig:introfig} can be understood in the low- and high-density limits as follows.
When $gn\lesssim \hbar\omc$, the condensate is restricted to the LLL and shows a Gaussian transverse density profile with a $1/e$ radius of $\lb$~\cite{Fletcher:2019,SI}. 
A Bogoliubov analysis around this state generically results in a Hamiltonian of the form~\cite{Sinha:2005}
\begin{equation}
\hat{H}_\textrm{LLL}=\sum\limits_{k>0}A_k \left(\hat{a}^\dagger_k\hat{a}_k+\hat{a}^\dagger_{-k}\hat{a}_{-k}\right)+B_k \left(\hat{a}^\dagger_k \hat{a}^\dagger_{-k}+\hat{a}_k \hat{a}_{-k}\right),
\label{eqn:pairProdHam}
\end{equation}
where $\hat{a}_k$ is the annihilation operator for a particle with momentum $\hbar k$ along the $y$-direction. This Hamiltonian describes pairs of modes ${\pm k}$, with natural frequency $A_k/\hbar$ and coupled by a pair-creation operator of strength $B_k$ which corresponds to a two-mode squeezing interaction in the language of quantum optics~\cite{Loudon:1987}. 
In a non-rotating uniform condensate, ${A_k=\frac{\hbar^2 k^2}{2m}{+}gn}$ and ${B_k=gn}$~\cite{Fetter:1971} and hence pair-creation is always weaker than the mode energy, leading to stable excitations~\cite{Bogoliubov:1947}.
However, the effective magnetic field profoundly changes this picture. First, in the LLL there is no kinetic energy contribution to $A_k$. Second, as illustrated in Fig.~\ref{fig:introfig}c, the coupling between momentum and position means that states with ${k\neq0}$ have a reduced overlap with the condensate and a correspondingly lower interaction energy. 
One finds~\cite{Sinha:2005} ${A_k=gn\left[2\exp(-k^2\lb^2/2)-1\right]/\sqrt{2}}$ and ${B_k=gn\exp(-k^2\lb^2)/\sqrt{2}}$, and the resulting dispersion ${\varepsilon_k=\sqrt{|A_k|^2-|B_k|^2}}$ is shown in Fig.~\ref{fig:introfig}d.
The spectrum is imaginary for an entire range of wavevectors ${k>0}$ beyond the zero-energy Goldstone mode at ${k=0}$, indicating dynamical instability of the Goldstone branch and correlated exponential growth of $\pm k$ pairs of these modes. Their interference with the ${k=0}$ condensate results in a density modulation (see Fig.~\ref{fig:introfig}c). 
The fastest growth occurs at a wavevector ${\sim1/\lb}$ giving a spatial modulation wavelength ${\sim 2\pi}$ times the magnetic length. This mode eventually becomes macroscopically occupied, corresponding to condensation of magneto-rotons and yielding a density modulation contrast of order unity.
Crucially, since interactions provide the only energy scale in the LLL, the instability growth rate is determined purely by the interaction energy $gn$. 

In the high-density limit where ${gn\gg\hbar\omc}$, a hydrodynamic description neglecting quantum pressure is valid. In this regime, the condensate initially exhibits a Thomas-Fermi density profile ${n_{2\mathrm{D}}~{\propto} ~1{-}x^2/\rtf^2}$ where ${\rtf=\sqrt{\frac{2 gn}{ m \omc^2}} = \sqrt{\frac{2gn}{\hbar\omega_c}} \lb}$~\cite{Recati:2001,SI}.
The Coriolis force 
${2 m \textbf{v}\times\bm{\Omega}}$ on each fluid element resulting from the shear flow ${\textbf{v}=(0,-\omega_c x)}$ perfectly balances the local gradient of mean-field energy, resulting in an inhomogeneous equilibrium density despite the absence of any scalar potential.
Our hydrodynamic stability analysis about this equilibrium state reveals a dynamical snaking instability of the cloud~\cite{SI}, in analogy to the Kelvin-Helmholtz instability of counterflow in fluid layers~\cite{Chandrasekhar:1961,land87fluid}, and the diocotron instability of charged plasmas and electron beams~\cite{Davidson1991,Cerfon:2016}.
The absence of quantum pressure means that the Thomas-Fermi radius and cyclotron frequency provide the only lengthscale and rate. Within the hydrodynamic analysis the instability develops at a wavevector set by the condensate width, as in the LLL, but at a density-independent rate $\propto \omega_c$ in striking qualitative contrast to the growth rate in the LLL.

\begin{figure}[t]    \centering
   \includegraphics[width=\columnwidth]{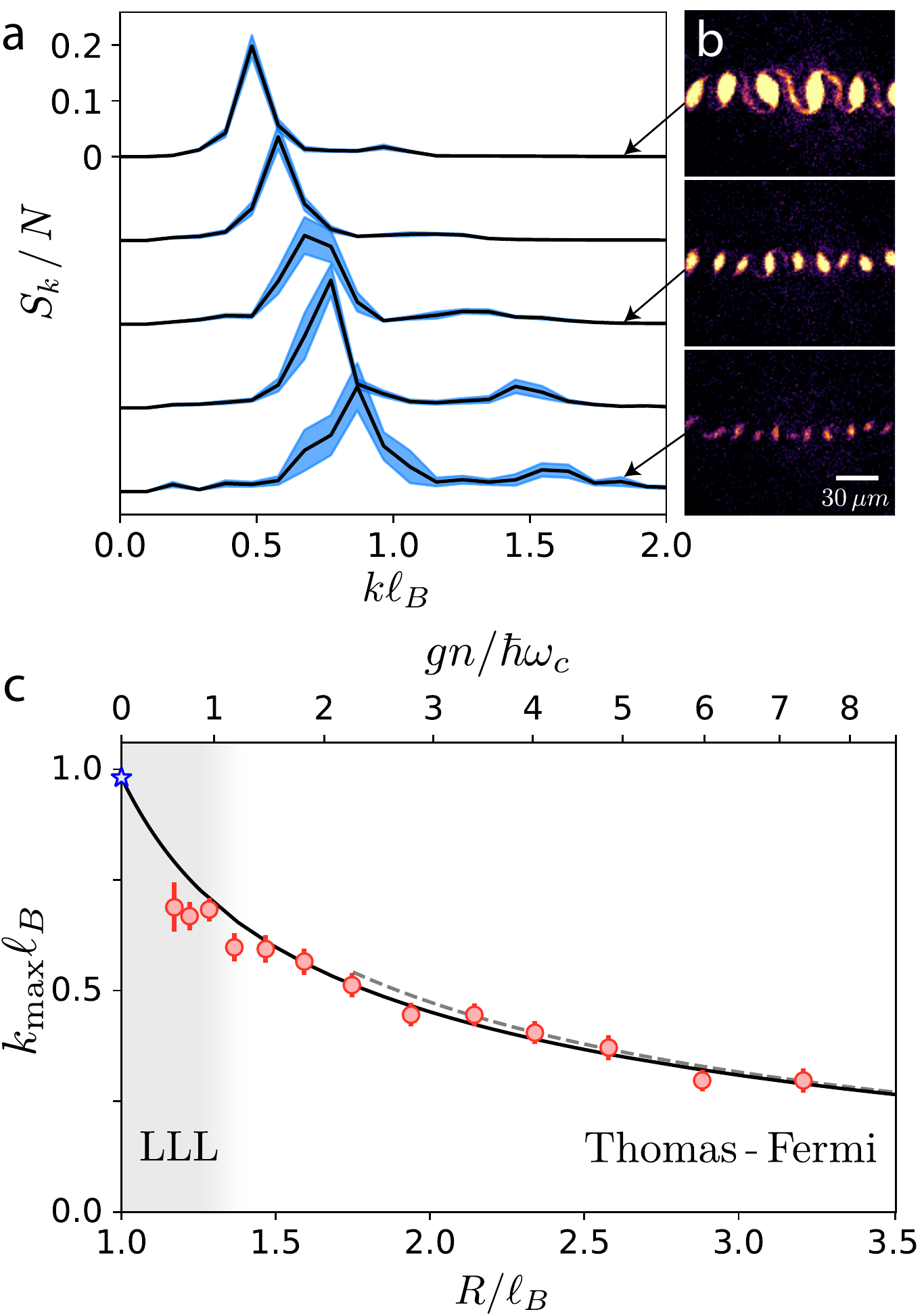} 
   \caption{
   \textbf{Structure factor and lengthscale of the emergent crystal.}
   \textbf{(a)}~The static structure factor, $S_k$, measured once the density modulation has reached steady-state for condensates with initial widths 
   ${R/\lb=2.58}$, $1.75$, $1.59$, $1.28$, and $1.22$ (top to bottom).
   The prominent peak reflects the periodic modulation of the cloud density.
   \textbf{(b)}~Corresponding images of the steady-state crystal, illustrating the decrease in the modulation lengthscale with falling condensate density.
   \textbf{(c)}~Dependence of the dominant modulation wavevector, $\ks$, on the cloud width, $R/\lb$. The LLL and hydrodynamic results are indicated by the star and dashed line respectively (see text). The solid line shows the prediction of our Bogoliubov analysis~\cite{SI}, which shows excellent agreement with our data with no free parameters. 
   }
   \label{fig:structureFactor}
\end{figure}

From these arguments, for all condensate densities we anticipate an emergent density modulation with a lengthscale set by the width of the initial cloud. For a quantitative analysis, from our experimental images (see Fig.~\ref{fig:introfig}e) we obtain the static structure factor ${S_k\equiv |n_k|^2/N}$, where ${n_k=\int\textrm{d}y\,n_\textrm{1D}(y)\,e^{-iky}}$ is the Fourier transform of the one-dimensional number density $n_\textrm{1D}(y)$~\cite{Hung:2011} and ${N=\int\textrm{d}y\,n_\textrm{1D}(y)}$. 
In Fig.~\ref{fig:structureFactor}a we show examples of $S_k$ obtained once the density modulation has fully developed, which show a well-defined peak at a wavevector $\ks$. We attribute the much smaller secondary peak at $ 2\ks$ to the contiguous traces of condensate linking adjacent droplets. In Fig.~2c we show $\ks$ as a function of the condensate density, which is parameterized by the ratio $R/\lb$ where $R$ is the full-width-at-half-maximum of the initial cloud divided by $2\sqrt{\log2}$. This normalization is chosen such that ${R/\lb\rightarrow1}$ for vanishing $gn$, while in the high-density limit ${R/\lb = \sqrt{gn/(\hbar \omc \log{2})}}$. At all densities, we indeed find an instability lengthscale of order the cloud width, ${\ks\sim 1/R}$. The star indicates the LLL prediction ${\ks=0.98/\lb}$ and the dashed line shows the hydrodynamic result ${\ks=0.95 / R}$~\cite{SI} neglecting quantum pressure. The solid line presents $\ks$ that we obtain from a numerical solution of the Bogoliubov equations~\cite{SI}, showing excellent agreement with the data without any free parameters. 

\begin{figure}[t] 
   \centering
   \includegraphics[width=\columnwidth]{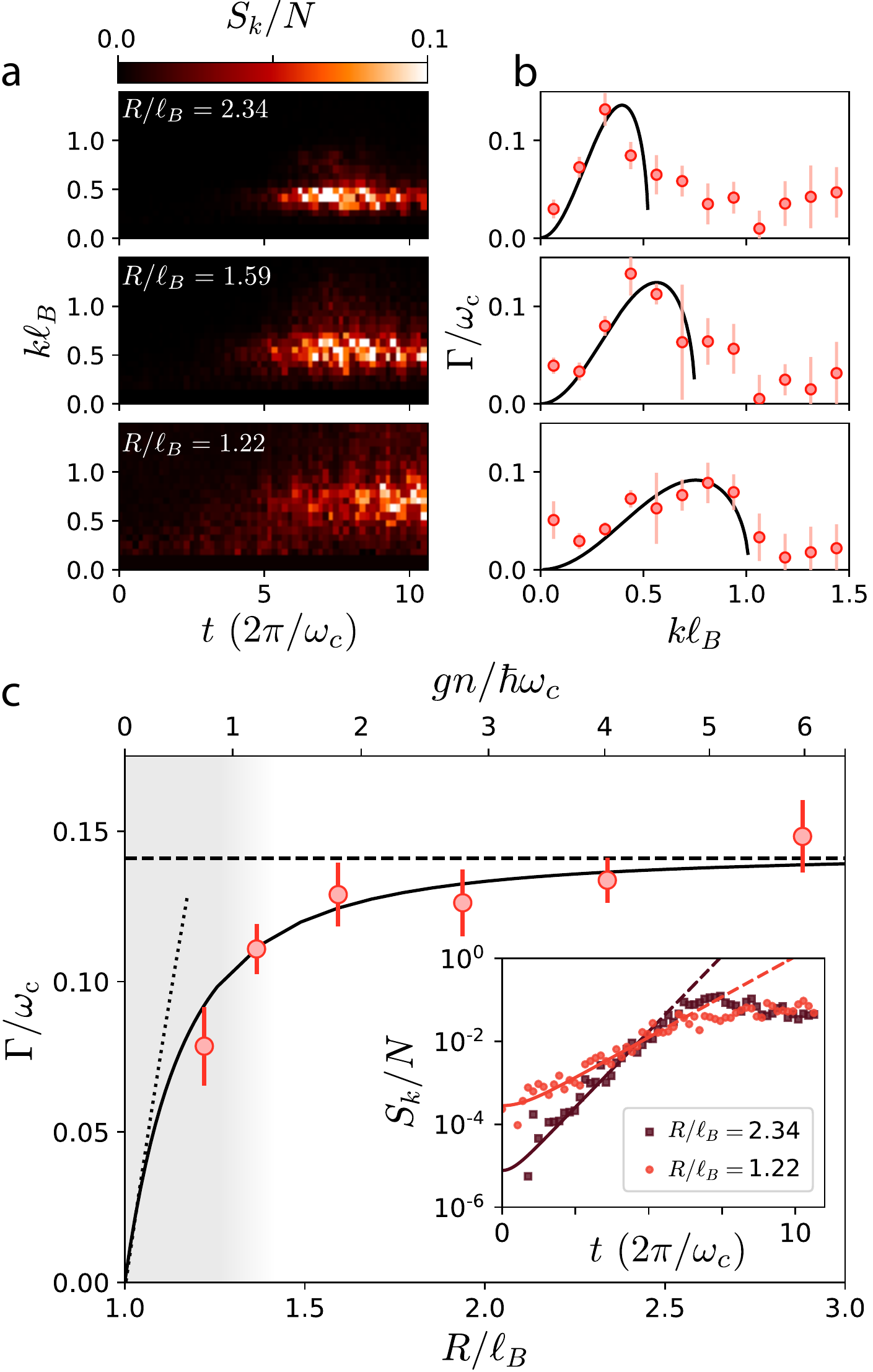} 
   \caption{
   \textbf{Instability growth dynamics.}
   \textbf{(a)}~The temporal evolution of the static structure factor, $S_k(t)$, for condensates with different initial widths which reveals a density-dependence of both the modulation lengthscale and growth rate. 
   \textbf{(b)}~The measured instability growth rate, $\Gamma$, as a function of wavevector. The solid line shows the rate obtained from our Bogoliubov analysis~\cite{SI} and captures the data well with no free parameters.
   \textbf{(c)}~The instability growth rate at the dominant unstable wavevector as a function of the condensate width. At high densities we find good agreement with the density-independent hydrodynamic rate $\Gamma=0.14\,\omc$ (dashed line). As the density falls, we observe a crossover to the LLL scaling $\Gamma= 0.21gn/\hbar$ (dotted line). Solid line: Bogoliubov analysis~\cite{SI}. The inset shows $S_k(t)$ at $\ks$ for condensates in the hydrodynamic regime (dark red) and the LLL (light red), along with the corresponding fits used to extract the rate (see text). 
   }
   \label{fig:growthRates}
\end{figure}

While the cloud width sets the instability lengthscale in both the LLL and hydrodynamic regimes, the growth rate shows qualitatively different behaviour. In Fig.~\ref{fig:growthRates}a, we show $S_k$ as a function of time for several different condensate densities. 
In addition to the decrease in the instability lengthscale at lower densities, we observe a concurrent reduction of the growth rate.
At each wavevector we fit the time-evolution of the structure factor with the theoretically expected function ${S_k(t)=A\cosh(2\Gamma t)}$~\cite{SI}, and extract the instability growth rate $\Gamma(k)$. This is reported in Fig.~\ref{fig:growthRates}b, along with the imaginary component of the corresponding Bogoliubov spectrum which shows good agreement without any free parameters. 
We note that the experimental data also reveal some growth in $S_k$ at higher wavevectors than the unstable region predicted by the linear Bogoliubov analysis. We attribute this to non-linear effects, and have performed numerical simulations of the Gross-Pitaevskii (GP) equation, finding that these exhibit the same behaviour~\cite{SI}.

We capture the typical crystallization rate corresponding to a particular condensate density by the growth rate of the dominant instability, $\Gamma(\ks)$, and in Fig.~\ref{fig:growthRates}c plot this as a function of $R/\lb$. 
When ${R/\lb\gg1}$ the rate is density-independent and consistent with our hydrodynamic result ${\Gamma=0.14\,\omc}$, shown by the dashed line. 
However, for lower interaction energies the gas enters the LLL where $gn$ provides the only energy scale. We observe a concurrent slowing down of the instability, and the data approach the LLL prediction ${\Gamma=0.21gn/\hbar}$ indicated by a dotted line.
At all densities, the data show good agreement with the rate obtained from our Bogoliubov analysis, reported as the solid line. 

After its initial hyperbolic growth, $S_k$ reaches a steady-state as shown in the inset of Fig.~\ref{fig:growthRates}c.
The emergent crystal is long-lived, with each droplet stabilized by a balance of the outward mean-field pressure and an inwards Coriolis force. This arises from the circulating flow within each droplet which is imposed by the gauge field, and is evident from vortices intersecting adjacent droplets (see Fig.~\ref{fig:introfig}e). The counterflow speed at the interface of two droplets of radius $R$ is ${\sim\omega_c R}$, giving a gradient of ${m\omega_c R/\hbar}$ in the relative phase and a vortex spacing of $2\pi \lb^2/R$. Adjacent droplets are therefore separated by ${\sim(R/\lb)^2}$ vortices. In the limit of classical hydrodynamics this number is large and the quantization of circulation is irrelevant, while in the LLL adjacent droplets are separated by a single vortex~\cite{Sinha:2005}.

\begin{figure}[t] 
   \centering
   \includegraphics[width=.97\columnwidth]{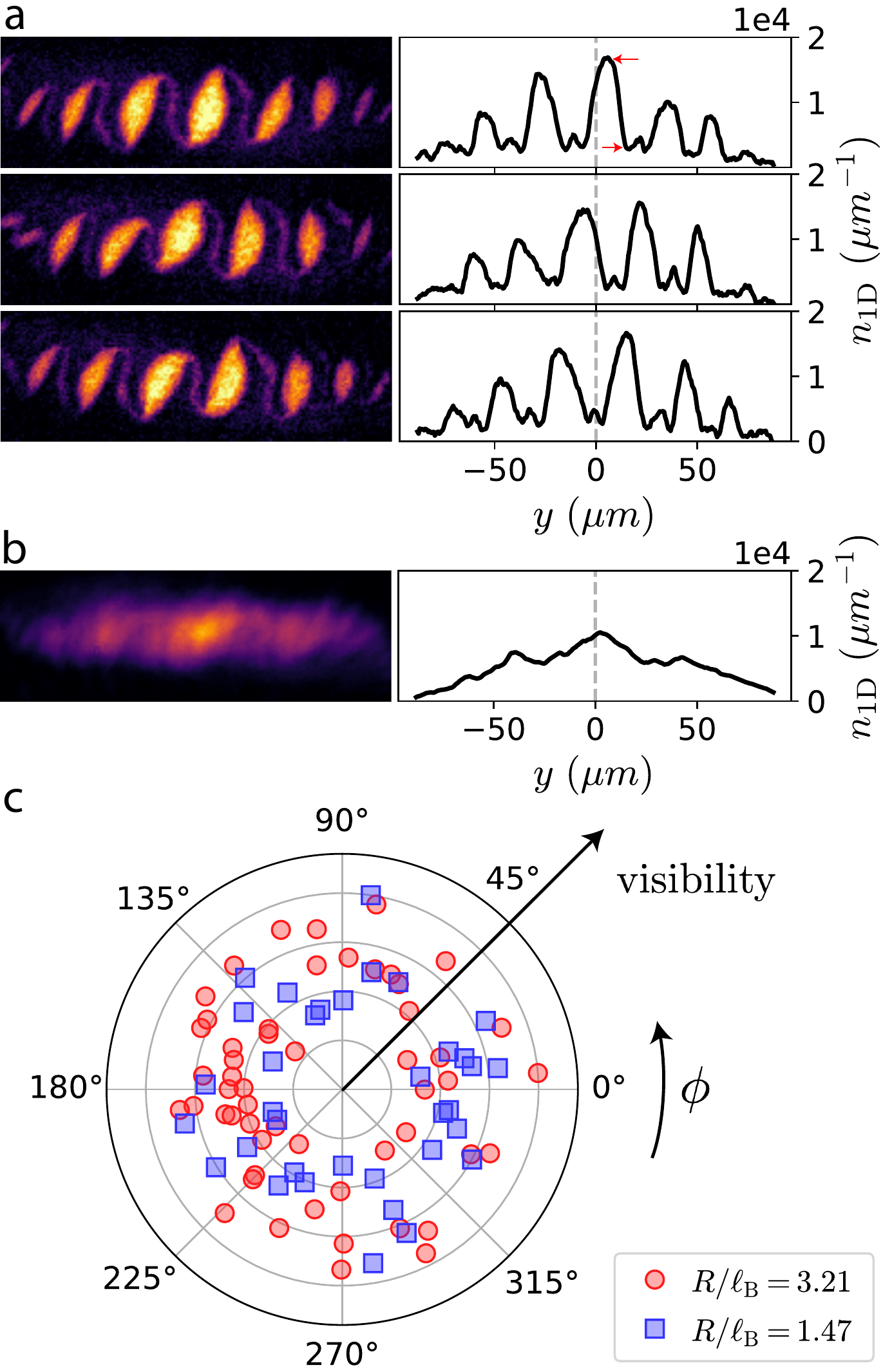} 
   \caption{
   \textbf{Spontaneous breaking of translational symmetry.}
   \textbf{(a)}~Images of the emergent crystal in three experimental iterations, along with the integrated one-dimensional density profiles $n_\textrm{1D}(y)$. The vertical dashed line shows the position of the centre-of-mass of the cloud, relative to which the modulation phase is random.
   \textbf{(b)}~An image of the cloud averaged over $60$ iterations, in which the density modulation is no longer visible.
   \textbf{(c)}~The phase, $\phi$, and visibility of the density modulation measured for multiple iterations of the experiment, for two different initial condensate densities. The phase is randomly distributed between $0$ and $2\pi$, indicating spontaneous breaking of the initial translational symmetry of the cloud. The visibility is defined as the contrast of the density modulation in the center of the cloud (shown as red arrows in  \textbf{(a)}), relative to the largest visibility for each cloud width. The visibility appears largely independent of the phase chosen by the modulation.
   }
   \label{fig:randomPhase}
\end{figure}

While the dynamical instability drives the growth of a density modulation, the initial seeding of the unstable mode must arise from thermal or quantum fluctuations in the gas density at ${t=0}$~\cite{SI}. Since the phase of these fluctuations is random, this results in spontaneous breaking of the initial translational symmetry of the condensate. In Fig.~\ref{fig:randomPhase} we show the phase and visibility of the density modulation observed in different iterations of our experiment. To account for small fluctuations in the overall cloud position, we fit the one-dimensional density profile with a sinusoidal function modulated by a Gaussian envelope, and obtain the modulation phase $\phi$ relative to the centre-of-mass of the cloud. %
At all densities we find that the phase is uncorrelated between different experimental realizations, indicating spontaneous breaking of the initial translational symmetry. 

The emergent crystallization observed here offers a pristine example of collective physics arising purely from the interplay of interparticle interactions and a gauge field. 
A natural immediate direction concerns the Goldstone mode associated with the spontaneous breaking of translational symmetry, corresponding to magneto-phonons in the droplet array~\cite{Jang2017Wigner}. 
This would be a remarkable instance of a propagating mode arising intrinsically from interactions, in the absence of any single-particle dynamics. 
While the densities in our experiment correspond to tens of atoms per flux quantum, our protocol can be straightforwardly extended to prepare clouds of lower filling fractions, which are expected to host beyond-mean-field, strongly correlated bosonic quantum Hall states~\cite{Cooper:2001,Ho:2001,Oktel:2004,Sinha:2005,aftalion:2009,Chen:2012,Senthil:2013,Vishwanath:2013}.

\begin{center}
    \textbf{Methods}
\end{center}
\paragraph{Preparation of Landau gauge condensates}
We prepare condensates occupying a single Landau gauge wavefunction using the geometric squeezing protocol described in~\cite{Fletcher:2019}.
We begin with a condensate of $8.1(1)\times10^5$ atoms of $^{23}$Na in an elliptical time-orbiting-potential (TOP) trap~\cite{Petrich:1995}, with a rms radial frequency ${\omega=2\pi\times88.6(1)}~$Hz, ellipticity $0.125(4)$, and axial frequency $2.8\,\omega$. 
We then rotate the ellipticity of the trap, ramping the rotation frequency from zero to $\omega$. In the rotating frame, atoms experience both a synthetic magnetic field and a scalar saddle potential.
Isopotential flow on this saddle, in analogy to the $\vec{E}\times\vec{B}$ Hall drift of electromagnetism, leads to elongation and contraction of the condensate along orthogonal directions and effecting unitary squeezing of the atomic density distribution~\cite{Fletcher:2019}.
We then turn off the saddle potential by setting the trap ellipticity to zero, which halts the outward flow of atoms.
This results in an equilibrium, quasi-translationally-invariant condensate freely rotating at $\omega$, which we allow to evolve for a variable time $t$.
Finally, we obtain an absorption image of the {in situ} density distribution. 

\paragraph{Imaging setup}
Our imaging resolution is sufficient to observe vortices {in situ} with a contrast of $\sim 60\%$~\cite{Fletcher:2019}. These have a characteristic size set by the healing length, which is $\sim 300~$nm in our system. This is significantly smaller than the quantum-mechanical ground state size of cyclotron orbits, set by the rotational analog of the magnetic length, ${\lb=\sqrt{\hbar/(2m\omega)}=1.6~\mu}$m.

\paragraph{Coupling constant}
Given interaction energies close to the LLL, the axial motion at frequency $2.8\omega$ is predominantly in its ground state. The coupling constant is then ${g=\frac{4\pi\hbar^2a}{m}{\mathrm \int } d z\left|\phi(z)\right|^4}$, where $a$ is the three-dimensional $s$-wave scattering length, and $\phi(z)$ is the axial wavefunction with normalization ${\int d z\,|\phi(z)|^2=1}$.

\newpage
\begin{center}
    \textbf{Acknowledgments}
\end{center}
\vskip -8pt
We thank T. Pfau and his research group, T. Senthil, T. Simula, and W. Zwerger for stimulating discussions.
This work was supported by
the National Science Foundation (Center for Ultracold Atoms and Grant No. PHY-2012110), 
Air Force Office of Scientific Research (FA9550-16-1-0324 and
MURI Quantum Phases of Matter FA9550-14-1-0035), 
Office of Naval Research (N00014-17-1-2257), 
the DARPA A-PhI program through ARO grant W911NF-19-1-0511, 
and the Vannevar Bush Faculty Fellowship.
A.S. acknowledges support from the NSF GRFP.
M.Z. acknowledges funding from the Alexander von Humboldt Foundation.

\bibliography{Crystallization_Mukherjee.bib, Crystallization_Mukherjee_More.bib}

\clearpage


\section*{Supplementary material}

# Crystallization of Bosonic Quantum Hall States
# Supplementary Information

Biswaroop Mukherjee, Airlia Shaffer, Parth B. Patel, Zhenjie Yan,
Cedric C. Wilson, Valentin Crépel, Richard J. Fletcher, Martin Zwierlein

*MIT-Harvard Center for Ultracold Atoms, Research Laboratory of Electronics, and Department of Physics, Massachusetts Institute of Technology, Cambridge, Massachusetts 02139, USA*

## GROSS-PITAEVSKII SIMULATION

To provide insight into the crystallization dynamics beyond what can be captured in the linear stability analysis presented below, we perform a numerical simulation of our experiment based upon the Gross-Pitaevskii (GP) equation. Within a single-mode approximation, the condensate wavefunction $\psi(\mathbf{r}, t)$ evolves in the rotating frame as,

$$i\hbar \frac{\partial}{\partial t} \psi(\mathbf{r}, t) = \left[ \frac{-\hbar^2 \nabla^2}{2m} + V(\mathbf{r}) + g|\psi(\mathbf{r}, t)|^2 - \mathbf{\Omega} \cdot \mathbf{L} \right] \psi(\mathbf{r}, t). \quad (1)$$

Here, $m$ denotes the atomic mass of $^{23}$Na, $g = \sqrt{8\pi} \frac{\hbar^2 a_s}{m l_z}$ is the two-dimensional mean-field coupling constant, $a_s = 3.3$ nm is the scattering length, $l_z = \sqrt{\frac{\hbar}{m \omega_z}}$ is the harmonic oscillator length of the axial trap, $\omega_z = 2.8\,\omega$ is the trap frequency in the z-direction, $\omega = 2\pi \times 88.6$ Hz is the rms radial trap frequency, $\mathbf{\Omega} = \Omega(t)\hat{z}$ is the angular velocity, $\mathbf{L}$ is the angular momentum operator, and $V$ is a complex scalar potential. The real part $\mathrm{Re}(V) = \frac{1}{2} m \omega^2 [(1+\varepsilon)x^2 + (1-\varepsilon)y^2]$ is the radial trapping potential with ellipticity $\varepsilon$, while the imaginary part $\mathrm{Im}(V) \propto 1 + \mathrm{erf}[(r - R_\infty)/\sigma)]$ serves as an absorbing circular boundary. The absorbing radius $R_\infty$ is chosen to be much larger than the transverse size of the condensate, and we use a wall thickness $\sigma = R_\infty/10$. We implement the evolution of Eq. (1) on a square grid using the time-splitting spectral method [1] and accelerate the simulation by performing the bulk of the computation on a graphics processing unit (GPU).

The simulated experimental sequence is identical to the experiment. We first perform geometric squeezing of an initially circular condensate [2], before setting the trap ellipticity $\varepsilon \to 0$ after which the condensate evolves freely for a time $t$ in the rotating frame. We find that without the explicit addition of noise, the condensate does not exhibit any instability except near the boundaries, due to residual edge effects not mitigated by the absorbing potential (see Fig. 1a). On the other hand, seeding of the dynamical instability by the addition of gaussian phase noise at time $t = 0$ results in a very similar simulated evolution (Fig. 1b) compared to the experiment (Fig. 1c).

We perform an identical analysis procedure as in the experiment (see main text) on the simulated density profiles in order to obtain the structure factor $S_k(t)$, shown in Figs. 1d-e, and the instability growth rate shown in Fig. 1f. The red points show the experimental instability growth rate as a function of wavevector $k$, while the black line shows the prediction of our Bogoliubov analysis below. For comparison, the blue line shows the rate extracted from the simulation, which captures the observed growth at higher wavevectors than the unstable range predicted by the Bogoliubov approach. This suggests that such growth can indeed be attributed to non-linear effects, which are not captured by the perturbative Bogoliubov approach. In addition to oscillations in $S_k$ at the cyclotron frequency $\omega_c$, a slower modulation is also observed. We attribute this oscillation to rotation of the individual droplets in the crystal.

In both experiment and simulation the emergent crystal is long-lived, persisting for $\omega_c t/(2\pi) > 20$. In the experiment the lifetime is only limited by the weak $\propto r^4$ anharmonicities in the trapping potential, leading to a slow S-shaped distortion of the linear crystal, similar to the Kerr effect on non-classical states in quantum optics.

### Vortex detection and phase profile:

In the rotating frame, each droplet exhibits an irrotational flow profile, driven by vortices surrounding the droplets. These vortices are directly visible in the experimental density image, and can be used to reconstruct the phase profile of the crystal in the rotating frame (see Fig. 2(a, c, e)). The phase is determined by the locations of the vortices, which are assumed to each have a single unit of circulation $2\pi\hbar/m$. Most vortices are outside of the bulk of the condensate, making their detection challenging. Nevertheless, a numerical solution of the GP equation shows similarly located vortices (Fig. 2d), as well as a similar irrotational flow profile in the rotating frame (Fig. 2f).

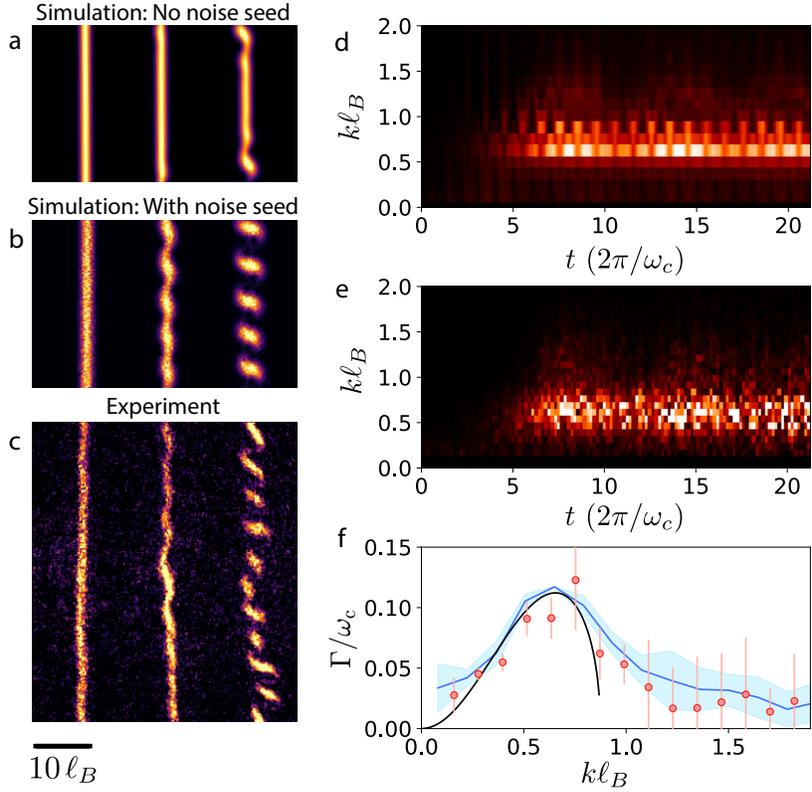

FIG. 1. Numerical GP simulation of the condensate evolution in the rotating frame. **(a-c)** Time evolution of the condensate density without the addition of noise (top), with added phase noise (middle), and in the experiment (bottom). The frames correspond to times $\omega_c t/(2\pi) = 0, 4,$ and 6. **(d-e)** Evolution of the structure factor $S_k(t)$ extracted from the simulation (d) and the experiment (e) which show good agreement. **(f)** The extracted instability growth rate as a function of wavevector $k$. The experimental measurements are shown by red points, and the Bogoliubov prediction by the black line. The blue line shows the result of the GP simulation, and the blue shading the statistical error reflecting variation across multiple iterations of the simulation. This model captures the experimentally measured growth at wavevectors above the instability region provided by the linear Bogoliubov description.

## BOGOLIUBOV STABILITY ANALYSIS OF LANDAU GAUGE CONDENSATES

A condensate prepared to have uniform phase in the Landau gauge - a "Landau gauge condensate" - is energetically unstable: an infinitely extended Bose gas has lower energy. However, in the absence of dissipation - as in the experiment - this does not itself lead to an instability. The question is whether the system is dynamically unstable, which would cause exponential growth of excitations and an isoenergetic transition into a new state. To search for dynamically unstable modes and to obtain their growth rate and spatial structure we perform a stability analysis of Landau gauge condensates via the Bogoliubov approach [3–5]. We expect the initial unmodulated condensate to be well-described within a single-mode framework, and expand the Hamiltonian in the Landau gauge, Eq. 2 of the main text, to second order in small fluctuations $\delta\hat{\psi}$ of the bosonic field $\hat{\Psi} = \psi_0 + \delta\hat{\psi}$ about the initial condensate wavefunction $\psi_0$ [3–5]. To ensure number conservation one employs the grand-canonical Hamiltonian $\hat{K} = \hat{H} - \mu\hat{N}$ with chemical potential $\mu$:

$$\hat{K} = \hat{H} - \mu\hat{N} \approx \text{const.} + \int d^2r\, \delta\hat{\psi}^\dagger \left(\hat{h} - \mu_0 + 2g\,|\psi_0|^2\right) \delta\hat{\psi} + \frac{1}{2}g \int d^2r \left(\psi_0^{*2}\delta\hat{\psi}\,\delta\hat{\psi} + \delta\hat{\psi}^\dagger\delta\hat{\psi}^\dagger\psi_0^2\right). \quad (2)$$

with the single-particle Hamiltonian in the Landau gauge $\hat{h} = \frac{\hat{p}_x^2}{2m} + \frac{1}{2}m\omega_c^2\left(\hat{x} - \frac{\hat{p}_y l_B^2}{\hbar}\right)^2$ and the constant term a function of $\psi_0$. Terms first order in $\delta\hat{\psi}$ vanish if $\psi_0$ obeys the stationary Gross-Pitaevskii (GP) equation $\hat{h}\psi_0 + g\,|\psi_0|^2\,\psi_0 = \mu_0\psi_0$. Since the initial wavefunction $\psi_0 = \psi_0(x)$ is translationally invariant along $y$, this reads

$$\left(-\frac{\hbar^2}{2m}\frac{d^2}{dx^2} + \frac{1}{2}m\omega_c^2 x^2 + g\,|\psi_0(x)|^2\right)\psi_0(x) = \mu_0\psi_0(x), \quad (3)$$

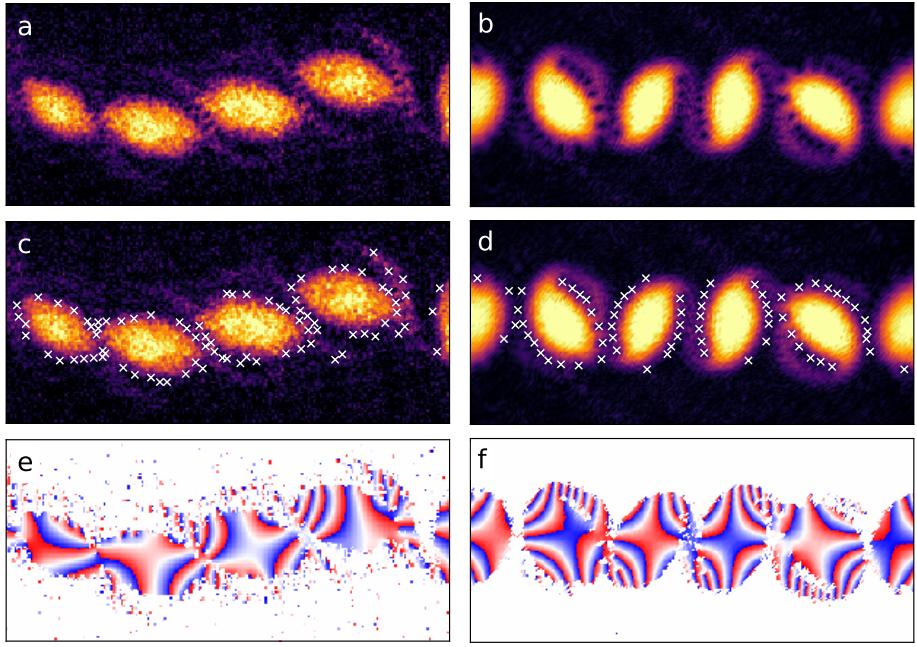

FIG. 2. Phase profile of the crystal (a, b) The density profiles of the crystals in the experiment (a) and GP simulation (b) appear to contain vortices which are marked in (c, d). (e) The phase of the macroscopic wavefunction can be inferred from the locations of the vortices in the experimental image. Note that additional contributions from undetected vortices may exist. (f) The simulated phase profile from a GP simulation shows a similar structure of irrotational flow within each segment of the crystal. In both (e) and (f), the phase shown is in the rotating frame.

which is formally equivalent to the GP equation of a Bose-Einstein condensate in a one-dimensional harmonic oscillator (h.o.) potential of frequency $\omega_c$. We solve Eq. 3 numerically via imaginary time evolution, choosing $\psi_0(x)$ to be real and normalized such that $\int \mathrm{d}x\, |\psi_0(x)|^2 = n_{1D}$, where $n_{1D}$ is the initial one-dimensional number density. A dimensionless quantity measuring the interaction energy, relative to the cyclotron level spacing $\hbar\omega_c$, is $\tilde{g} \equiv \frac{g n_{1D}}{l_B \hbar \omega_c}$. Near the lowest Landau level, the interaction is a small perturbation, resulting in the gaussian density $\psi_0^2(x) = \frac{n_{1D}}{\sqrt{\pi} l_B} e^{-x^2/l_B^2}$ and $\mu_0 \approx \frac{\hbar\omega_c}{2}\left(1 + \sqrt{\frac{2}{\pi}}\tilde{g}\right)$, close to the ground-state cyclotron energy. In the Thomas-Fermi regime, where the term $\hat{p}_x^2/2m$ can be neglected, one obtains $g\psi_0^2(x) = \mu_0 - \frac{1}{2}m\omega_c^2 x^2$ with $\mu_0 = \frac{1}{2}\hbar\omega_c \left(\frac{3}{2}\tilde{g}\right)^{2/3}$.

Translation invariance of $\psi_0(x)$ and $\hat{h}$ along $y$ allows expanding $\delta\hat{\psi} = \sum_k \frac{1}{\sqrt{L}} e^{iky} \hat{\phi}_k(x)$ into bosonic fields $\hat{\phi}_k(x)$ of well-defined $y$-momentum $\hbar k$, with $L$ the spatial extent of the system in the $y$-direction. The quadratic part of Eq. 2 then becomes, using matrix notation,

$$\hat{K}_2 = \frac{1}{2}\sum_k \int \mathrm{d}x \left(\hat{\phi}_k^\dagger\ \hat{\phi}_{-k}\right) \begin{pmatrix} \hat{h}_k - \mu_0 + 2g\psi_0^2 & g\psi_0^2 \\ g\psi_0^2 & \hat{h}_{-k} - \mu_0 + 2g\psi_0^2 \end{pmatrix} \begin{pmatrix} \hat{\phi}_k \\ \hat{\phi}_{-k}^\dagger \end{pmatrix}$$
$$= \frac{1}{2}\sum_k \langle \hat{\boldsymbol{\Phi}}_k | \hat{\mathbf{H}}_k | \hat{\boldsymbol{\Phi}}_k \rangle \qquad (4)$$

with $\hat{h}_k = \frac{\hat{p}_x^2}{2m} + \frac{1}{2}m\omega_c^2\left(x - kl_B^2\right)^2$, $\hat{\boldsymbol{\Phi}}_k = \left(\hat{\phi}_k\ \hat{\phi}_{-k}^\dagger\right)^T$, $\hat{\mathbf{H}}_k(x)$ the $2\times 2$ matrix operator of the first line in Eq. 4, and $\langle \mathbf{f}|\mathbf{g}\rangle = \int \mathrm{d}x\, \mathbf{f}^\dagger(x)\cdot\mathbf{g}(x)$ for vectors $\mathbf{f}, \mathbf{g}$. Momentum conservation along $y$ ensures that allowed scattering processes result in either the simultaneous creation or simultaneous annihilation of a pair of states with momenta $k$ and $-k$. Consequently the Hamiltonian only mixes a particle with y-momentum $k$ with a hole of y-momentum $-k$, as is explicit in the $2\times 2$ particle/hole matrix notation. The Bogoliubov Hamiltonian $\hat{\mathbf{H}}_k(x)$ in Eq. 4 is Hermitian, has only real eigenvalues (bounded from below by $-\mu_0$) and thus $\hat{K}_2$ has only real expectation values in any state. However, the time evolution of the bosonic field operators

$\hat{\phi}_k(x,t)$ evolving under the grand-canonical Hamiltonian $\hat{K}$ is given by $i\hbar\frac{\partial}{\partial t}\hat{\phi}_k = \left[\hat{\phi}_k, \hat{K}\right]$, and with the bosonic commutation relations $\left[\hat{\phi}_k(x), \hat{\phi}_{k'}^\dagger(x')\right] = \delta_{k,k'}\delta(x-x')$ we have

$$i\hbar\frac{\partial}{\partial t}\hat{\mathbf{\Phi}}_k = \begin{pmatrix} \hat{h}_k - \mu_0 + 2g\psi_0^2 & g\psi_0^2 \\ -g\psi_0^2 & -(\hat{h}_{-k} - \mu_0 + 2g\psi_0^2) \end{pmatrix} \hat{\mathbf{\Phi}}_k = \eta\hat{\mathbf{H}}_k\hat{\mathbf{\Phi}}_k \quad (5)$$

with $\eta = \begin{pmatrix} 1 & 0 \\ 0 & -1 \end{pmatrix}$ acting in particle-hole space [6, 7]. The evolution of the field operators is thus governed by an operator $\eta\hat{\mathbf{H}}_k$ that is in general *non*-Hermitian and can thus feature complex eigenvalues, leading to exponential growth of fluctuations - the system features dynamical instabilities [8].

### Symmetries, eigenvectors and eigenvalues of $\eta\hat{\mathbf{H}}_k$

For each $k$, the Hamiltonian matrix $\hat{\mathbf{H}}_k$ is real, $\hat{\mathbf{H}}_k^* = \hat{\mathbf{H}}_k$, and symmetric under simultaneous reflection of space $\hat{R}$ (i.e. $\hat{R}x\hat{R} = -x$) and exchange of particles and holes, i.e. $\hat{\mathbf{H}}_k = \gamma\hat{R}\hat{\mathbf{H}}_k\hat{R}\gamma$ with $\gamma = \begin{pmatrix} 0 & 1 \\ 1 & 0 \end{pmatrix}$ exchanging particles and holes. It follows that given an eigenvector $\mathbf{V}_{k,n}$ of $\eta\hat{\mathbf{H}}_k$ with eigenvalue $\epsilon_{k,n}$, the vector $\mathbf{V}_{k,n}^*$ is also an eigenvector with eigenvalue $\epsilon_{k,n}^*$, and $\gamma\hat{R}\mathbf{V}_{k,n}$ and $\gamma\hat{R}\mathbf{V}_{k,n}^*$ are eigenvectors with eigenvalues $-\epsilon_{k,n}$ and $-\epsilon_{k,n}^*$, respectively. The latter follows from $\gamma\eta = -\eta\gamma$. We also note that $\hat{\mathbf{H}}_{-k} = \hat{R}\hat{\mathbf{H}}_k\hat{R}$ implying that $\hat{R}\mathbf{V}_{k,n} \equiv \mathbf{V}_{-k,n}$ is eigenvector of $\eta\hat{\mathbf{H}}_{-k}$ with eigenvalue $\epsilon_{-k,n} = \epsilon_{k,n}$. In general, the four values $\epsilon_{k,n}$, $\epsilon_{k,n}^*$, $-\epsilon_{k,n}$ and $-\epsilon_{k,n}^*$ are all different, implying an oscillatory evolution of exponentially increasing and decreasing amplitudes. The instability studied in the present work concerns the mode of lowest $|\epsilon_{k,n}|$ for given $k$, the Goldstone branch which we label by $n=0$. It is smoothly connected to the Goldstone mode at $k=0$ of zero frequency, $\epsilon_{0,0}=0$, that reflects the free choice of the overall phase of the condensate, i.e. its $U(1)$ symmetry. An associated second mode with zero eigenvalue of $(\eta\hat{\mathbf{H}}_0)^2$ describes the global phase fluctuations [9]. The next excited mode, $n=1$, is correlated near $k=0$ with the cyclotron oscillation. At $k=0$ the $n=1$ mode lies precisely at the cyclotron energy $\epsilon_{0,1} = \hbar\omega_c$, according to Kohn's theorem [10, 11]. The modes at $k>0$ of the Goldstone branch, connecting to the Goldstone density and phase modes at $k=0$, are thus well separated from any other excitations, so that this branch is described by only two, not four, distinct eigenvalues. This implies that either $\epsilon_{k,0} = \epsilon_{k,0}^*$, i.e. one has two real eigenvalues $\pm\epsilon_{k,0}$, or $\epsilon_{k,0} = -\epsilon_{k,0}^*$, i.e. one has two purely imaginary eigenvalues $\pm\epsilon_{k,0}$. For excitations of non-rotating condensates in their ground state, only the first case occurs and corresponds to the usual Bogoliubov phonon excitations. Here, instead, we find, in an entire range of momenta between $k=0$ and a maximum $k=k_c$, the case of purely imaginary frequencies $\epsilon_{k,0}$, corresponding to the exponential growth of correlated excitations at $\pm k$ that causes the "snake-like" dynamical instability. Results of the numerical solution of $\eta\hat{\mathbf{H}}_k\mathbf{V}_{k,n} = \epsilon_{k,n}\mathbf{V}_{k,n}$ are shown in Fig. 3, from deep in the lowest Landau level ($\tilde{g} \lesssim 1$) to the Thomas-Fermi regime ($\tilde{g} \gg 1$).

### Relation between eigenvalues of $\hat{\mathbf{H}}_k$ and $\eta\hat{\mathbf{H}}_k$

The difference between stable and dynamically unstable excitations is analogous to the difference between the stable motion in a harmonic oscillator potential and the unstable motion of a particle in an inverted harmonic oscillator. The correspondence becomes explicit if we introduce the Hermitian operators $\hat{Q}_k(x) = \left(\hat{\phi}_k(x) + \hat{\phi}_k^\dagger(x)\right)/\sqrt{2}$ and $\hat{P}_k(x) = -i\left(\hat{\phi}_k(x) - \hat{\phi}_k^\dagger(x)\right)/\sqrt{2}$ obeying $\left[\hat{Q}_k(x), \hat{P}_{k'}(x')\right] = i\delta_{k,k'}\delta(x-x')$. They are related to the density and current fluctuations of the condensate, as the density operator is

$$\hat{n}(x,y) = \hat{\Psi}^\dagger\hat{\Psi} \approx |\psi_0|^2 + \psi_0(\delta\hat{\psi} + \delta\hat{\psi}^\dagger) = |\psi_0|^2 + \psi_0\sqrt{\frac{2}{L}}\sum_k \cos(ky)\hat{Q}_k(x) - \sin(ky)\hat{P}_k(x)$$

and $\hat{\mathbf{j}} \approx |\psi_0|^2 \nabla\hat{\Theta}$ with the linear fluctuation part of the velocity potential operator [4]

$$\hat{\Theta} = \frac{\hbar}{2mi\psi_0}\left(\delta\hat{\psi} - \delta\hat{\psi}^\dagger\right) = \frac{\hbar}{2m\psi_0}\sqrt{\frac{2}{L}}\sum_k \cos(ky)\hat{P}_k(x) + \sin(ky)\hat{Q}_k(x)$$

In terms of $\hat{Q}_k$ and $\hat{P}_k$, we have $\hat{K}_2 = \frac{1}{4}\sum_k \langle\hat{\mathbf{Q}}_k|\hat{\mathbf{H}}_k|\hat{\mathbf{Q}}_k\rangle + \langle\hat{\mathbf{P}}_k|\eta\hat{\mathbf{H}}_k\eta|\hat{\mathbf{P}}_k\rangle$ with $\hat{\mathbf{Q}}_k^T = (\hat{Q}_k\ \hat{Q}_{-k})$ and $\hat{\mathbf{P}}_k^T = (\hat{P}_k\ \hat{P}_{-k})$. So we can think of $\hat{\mathbf{H}}_k$ as representing the matrix of "spring constants" and $\eta\hat{\mathbf{H}}_k\eta$ the matrix of "inverse masses" in the oscillator

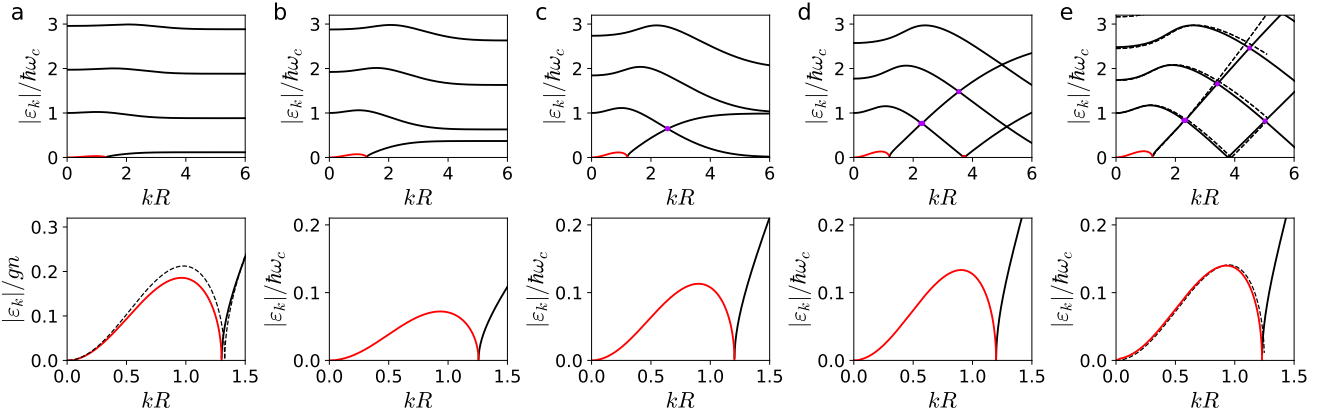

FIG. 3. Bogoliubov spectra from the Lowest Landau Level to the Thomas-Fermi regime. The interaction parameters $\tilde{g} \equiv \frac{gn_{1D}}{l_B \hbar \omega_c}$ for (a-e) are 0.3, 1, 3, 10, 40, corresponding to $R/l_B = 1.04, 1.14, 1.39, 2.01, 3.28$, capturing the evolution from the flat spectra deep in the LLL to the more intricate crossings in the Thomas-Fermi regime. The bottom panel shows a zoomed-in region focusing on the unstable Goldstone branch (red). In the bottom panel of (a), the growth rate/excitation frequencies are normalized by $gn$, with $n = |\psi_0(0)|^2 = n_{1D}/(\sqrt{\pi} l_B)$ the central 2D density in the LLL. The dashed line in this panel shows the result in the deep LLL limit using [12]. In the Thomas-Fermi regime, the dashed line in e) shows the result of the hydrodynamic calculation (see text). Note the existence of complex eigenvalues with $\text{Re}(\epsilon_k) \neq 0$ near the curve crossings, indicated by purple lines.

analogy. Dynamical instabilities can arise in an oscillator when either a spring constant becomes negative, while the mass remains positive, or vice versa.

The time evolution $\frac{d}{dt}\hat{\mathbf{Q}}_k = \left[\hat{\mathbf{Q}}_k, \hat{K}_2\right] = \eta \hat{\mathbf{H}}_k \eta \hat{\mathbf{P}}_k$ and $\frac{d}{dt}\hat{\mathbf{P}}_k = -\hat{\mathbf{H}}_k \hat{\mathbf{Q}}$ yields $\frac{d^2}{dt^2}\hat{\mathbf{Q}}_k = -\eta \hat{\mathbf{H}}_k \eta \hat{\mathbf{H}}_k \hat{\mathbf{Q}}_k \stackrel{!}{=} -\epsilon_k^2 \hat{\mathbf{Q}}_k$ showing that eigenfrequencies of the motion correspond indeed to the eigenvalues of the operator $\eta \hat{\mathbf{H}}_k$. The Hermitian operators $\hat{\mathbf{H}}_k$ and $\eta \hat{\mathbf{H}}_k \eta$ share their (real) eigenvalues, and if $\mathbf{U}_{k,n}$ is an eigenvector of $\hat{\mathbf{H}}_k$ of eigenvalue $E_{k,n}$, then $\eta \mathbf{U}_{k,n}$ is the eigenvector of $\eta \hat{\mathbf{H}}_k \eta$ with that same eigenvalue. With $\mathcal{O}_{k,nm} = \langle \mathbf{U}_{k,n} | \eta | \mathbf{U}_{k,m} \rangle = \int dx \, \mathbf{U}_{k,n}(x) \eta \mathbf{U}_{k,m}(x)$ the matrix effecting the basis change, which is symmetric and orthonormal (so $\mathcal{O}_k^2 = \mathbb{1}$), we have $\langle \mathbf{U}_{k,n} | \eta \hat{\mathbf{H}}_k \eta | \mathbf{U}_{k,m} \rangle = \sum_l \mathcal{O}_{k,nl} E_{k,l} \mathcal{O}_{k,lm} = (\mathcal{O}_k \mathcal{E}_k \mathcal{O}_k)_{nm}$, with $\mathcal{E}_k$ the diagonal matrix of eigenvalues of $\hat{\mathbf{H}}_k$ (the "spring constants"). The squared eigenfrequencies $\epsilon_k^2$ are thus eigenvalues of $\mathcal{O}_k \mathcal{E}_k \mathcal{O}_k \mathcal{E}_k$, and so the eigenfrequencies $\epsilon_k$ themselves are eigenvalues of $\mathcal{O}_k \mathcal{E}_k$, the matrix describing $\eta \hat{\mathbf{H}}_k$ in the basis of eigenvectors of $\hat{\mathbf{H}}_k$. Importantly, whenever a "spring constant" or "inverse mass" equals zero, i.e. one of the eigenvalues of $\hat{\mathbf{H}}_k$ equals zero, one eigenfrequency $\epsilon_k$ of $\eta \hat{\mathbf{H}}_k$ also equals zero. Regions in the variable $k$ featuring dynamical instabilities with purely imaginary eigenfrequency are thus bounded by values of $k$ where consecutive eigenvalues $E_k$ of $\hat{\mathbf{H}}_k$ equal zero. This is analogous to a harmonic oscillator slowing down and becoming dynamically unstable as its spring constant changes from positive to negative, followed by its mass diverging and changing sign to yield again a dynamically stable, but thermodynamically unstable, oscillator. A famous example of the latter situation is the magnetron motion in Penning traps [13].

Since $\hat{\mathbf{H}}_k$ commutes with simultaneous reflection and particle-hole exchange, i.e. with $\gamma \hat{R}$, eigenvectors of $\hat{\mathbf{H}}_k$ can be found as eigenvectors of $\gamma \hat{R}$ with eigenvalue $\sigma = +1$ or $-1$, which are states of the form $\mathbf{U}_\pm = (u(x), \pm u(-x))^T$, leading to the two eigenequations

$$(\hat{h}_k - \mu_0 + 2g\psi_0^2)u_{k,n\pm}(x) \pm g\psi_0^2 u_{k,n\pm}(-x) = E_{k,n\pm}u_{k,n\pm}(x)$$

The lowest energy for $\sigma = -1$ and $k=0$ is $E_{0,0-}=0$, for $u(x) = \psi_0(x)$, and $\mathbf{U}_{0,0-}(x) = (\psi_0(x), -\psi_0(x))^T/\sqrt{2}$ is the Goldstone mode. Since $\psi_0(x)$ is the ground state for a condensate trapped in a 1D harmonic oscillator, the Hamiltonian $\hat{\mathbf{H}}_0$, describing fluctuations that are translation invariant along $y$, is positive semi-definite, with eigenvalues $E_{0,n\pm}$ all positive or zero. For $k > 0$, the decreasing overlap of the eigenfunction $u_{k,0-}$ with the condensate centered at $x = 0$ causes the eigenvalue $E_{k,0-}$ to become negative, corresponding to the case of a negative spring constant. $\hat{K}_2$ then contains a term corresponding to an inverted oscillator potential, $\frac{1}{2}E_{k,0-}\hat{Q}_{k,0-}^2$ with $\hat{Q}_{k,n\sigma} \equiv \langle \mathbf{U}_{k,n\sigma} | \hat{\mathbf{Q}}_k \rangle$ the "position" operators, with the canonically conjugate "momentum" operators $\hat{P}_{k,n\sigma} \equiv \langle \mathbf{U}_{k,n\sigma} | \hat{\mathbf{P}}_k \rangle$ and commutation relations $\left[\hat{Q}_{k,n\sigma}, \hat{P}_{k,m\sigma'}\right] = i\delta_{nm}\delta_{\sigma\sigma'}$.

To obtain the matrix of inverse masses, we note that since $\eta$ anti-commutes with particle-hole exchange, $\eta\gamma = -\gamma\eta$, it only connects states of opposite $\gamma\hat{R}$ symmetry, so $\langle U_{k,n+}|\eta|U_{k,m+}\rangle = \langle U_{k,n-}|\eta|U_{k,m-}\rangle = 0$. With $\mathcal{O}_{k,nm}^{+-} \equiv \langle \mathbf{U}_{k,n+}|\eta|\mathbf{U}_{k,m-}\rangle =$

5

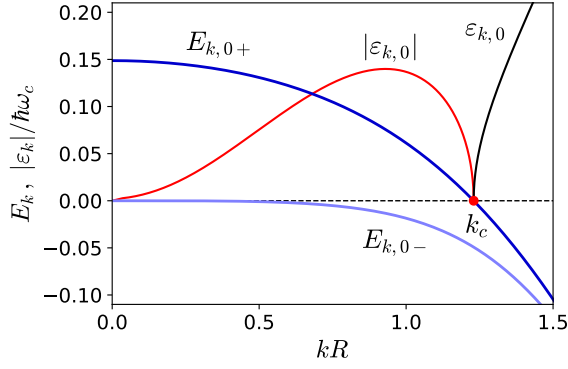

FIG. 4. Relation between Goldstone branch instability and eigenenergies of the Bogoliubov Hamiltonian $\hat{\mathbf{H}}_k$. The decrease of overlap with the condensate at non-zero $k$ renders the lowest eigenvalue $E_{k,0-}$ of $\hat{\mathbf{H}}_k$ negative, corresponding to an inverted oscillator potential (negative spring constant). The Goldstone branch is unstable, $\epsilon_{k,0}$ being purely imaginary, until the second eigenvalue $E_{k,0+}$ of $\hat{\mathbf{H}}_k$ becomes negative, corresponding to a negative mass oscillator and resulting in stable motion, with $\epsilon_{k,0}$ real.

$\int dx\, u_{k,n+}(x) u_{k,m-}(x) = \mathcal{O}^{-+}_{k,mn}$ we find for the "inverse mass" matrix $\langle \mathbf{U}_{k,n+}|\eta\hat{\mathbf{H}}_k\eta|\mathbf{U}_{k,m+}\rangle = \left(\mathcal{O}^{+-}_k \mathcal{E}_{k-} \mathcal{O}^{-+}_k\right)_{nm}$ and $\langle \mathbf{U}_{k,n-}|\eta\hat{\mathbf{H}}_k\eta|\mathbf{U}_{k,m-}\rangle = \left(\mathcal{O}^{-+}_k \mathcal{E}_{k+} \mathcal{O}^{+-}_k\right)_{nm}$ with $\mathcal{E}_{k\sigma}$ the diagonal matrix of eigenvalues $E_{k,\sigma}$. The squared Bogoliubov eigenfrequencies $\epsilon_k^2$ are thus eigenvalues of the matrix $\mathcal{O}^{-+}_k \mathcal{E}_{k+} \mathcal{O}^{+-}_k \mathcal{E}_{k-}$.

Fig. 4 shows the generic case. The branch $E_{k,0-}$, the lowest eigenvalue of $\hat{\mathbf{H}}_k$, is negative for any non-zero $k$, and the Goldstone branch $\epsilon_{k,0}$ (eigenvalue of $\eta\hat{\mathbf{H}}_k$) is correspondingly purely imaginary and thus dynamical unstable, until at $k=k_c$ the second branch $E_{k,0+}$ crosses zero, and $\epsilon_{k,0}$ becomes real. This is the situation of having both a negative mass and a negative spring constant, corresponding to dynamically stable motion. The point at $k=k_c$ is called an exceptional point in the theory of non-Hermitian physics [8]. $k_c$ is always on the order of the inverse cloud radius $k_c \sim 1/R$, and the maximum instability growth rate $|\epsilon_{k,0}|$ is also only slightly below $k_c$.

### Normal form of $\hat{K}_2$

An inverted harmonic oscillator Hamiltonian $H = \frac{p^2}{2m} - \frac{1}{2}\kappa q^2$ with negative spring constant $(-\kappa < 0)$, and mass $m > 0$ can be canonically transformed via $q' = \frac{1}{\sqrt{2m\Gamma}}p + \sqrt{\frac{\kappa}{2\Gamma}}q$ and $p' = \frac{1}{\sqrt{2m\Gamma}}p - \sqrt{\frac{\kappa}{2\Gamma}}q$ into $H = \frac{1}{2}\Gamma(q'p' + p'q')$, where $\Gamma = \sqrt{\kappa/m}$, with equations of motion $\frac{d}{dt}q' = \Gamma q'$ and $\frac{d}{dt}p' = -\Gamma p'$, generating squeezing of $p'$ and exponential growth of $q'$. In terms of bosonic operators $a = (q' + ip')/\sqrt{2}$ and $a^\dagger = (q' - ip')/\sqrt{2}$ with $[a, a^\dagger] = 1$ the Hamiltonian is of the squeezing form $H = \Gamma \frac{1}{2i}(aa - a^\dagger a^\dagger)$. Analogously we will find that $\hat{K}_2$, in the dynamically unstable region $0 < k < k_c$, will contain a term of the squeezing form, $\sum_k \Gamma_k \left(\hat{a}_k \hat{a}_{-k} + \hat{a}^\dagger_k \hat{a}^\dagger_{-k}\right)$ associated with the spontaneous pairwise creation of excitations at $\pm k$. Here, $\epsilon_k = i\Gamma_k$ with $\Gamma_k > 0$ is purely imaginary (we omit the index 0 in $\epsilon_{k,0}$ for simplicity), and we have, since $\epsilon_k = -\epsilon_k^*$, two instead of four associated eigenvectors of $\eta\hat{\mathbf{H}}_k$, labelled $\mathbf{V}_k$ and $\mathbf{W}_k = \mathbf{V}_k^*$, associated with the different eigenvalues $\epsilon_k$ and $\epsilon_k^* = -\epsilon_k$. To have $\epsilon_k = -\epsilon_k^*$, $\mathbf{V}_k$ and $\gamma\hat{R}\mathbf{V}_k^* = \gamma\mathbf{V}_{-k}^*$ must be linearly dependent, related by a complex phase $e^{i\theta}$. Choosing this phase corresponds to a particular choice of the spatial phase of the emergent crystal. We here set $\mathbf{V}_k = -\gamma\hat{R}\mathbf{V}_k^*$, and we then also have $\mathbf{W}_k = -\gamma\hat{R}\mathbf{W}_k^*$. From this follows that $\mathbf{V}_k = (u_k(x), -u_k^*(-x))^T$ and $\mathbf{W}_k = (u_k^*(x), -u_k(-x))^T$. We see $\langle\mathbf{V}_k|\eta|\mathbf{V}_k\rangle = 0 = \langle\mathbf{W}_k|\eta|\mathbf{W}_k\rangle$ but we can choose the normalization of $u_k(x)$ such that $\langle\mathbf{W}_k|\eta|\mathbf{V}_k\rangle = \int dx\, \left(u_k(x)^2 - u_k^*(x)^2\right) = i$. Then the action of $\hat{\mathbf{H}}_k$ on the subspace relevant to $\epsilon_k$ is

$$\hat{\mathbf{H}}_k = \Gamma_k\, \eta|\mathbf{V}_k\rangle\langle\mathbf{W}_k|\eta + \Gamma_k\, \eta|\mathbf{W}_k\rangle\langle\mathbf{V}_k|\eta + \ldots$$

with ... the part of $\hat{\mathbf{H}}_k$ corresponding to modes with $n > 0$. Inserting this in Eq. 4, and noting that $\mathbf{V}_{-k} = \hat{R}\mathbf{V}_k$ and so $u_{-k}(x) = u_k(-x)$ gives

$$\hat{K}_2 = \frac{1}{2}\sum_k \Gamma_k \left(\langle\hat{\boldsymbol{\Phi}}_k|\eta|\mathbf{V}_k\rangle\langle\mathbf{W}_k|\eta|\hat{\boldsymbol{\Phi}}_k\rangle + \langle\hat{\boldsymbol{\Phi}}_k|\eta|\mathbf{W}_k\rangle\langle\mathbf{V}_k|\eta|\hat{\boldsymbol{\Phi}}_k\rangle\right) + \ldots$$

$$= \frac{1}{2}\sum_k \Gamma_k \left(\hat{p}_k\hat{q}_{-k} + \hat{q}_k\hat{p}_{-k}\right) + \ldots$$

$$= \sum_{k>0}\Gamma_k \frac{1}{i}\left(\hat{a}_k\hat{a}_{-k} - \hat{a}_k^\dagger\hat{a}_{-k}^\dagger\right) + \ldots$$

where the dots denote contributions from stable modes of higher excitation energies $|\epsilon_{k,n}|$ and the $k=0$ Goldstone mode's "kinetic energy" term corresponding to free global phase diffusion [9], and we defined

$$\hat{p}_k \equiv \langle\hat{\boldsymbol{\Phi}}_k|\eta|\mathbf{V}_k\rangle = \int \mathrm{d}x\,\left(u_k(x)\phi_k^\dagger + u_k^*(-x)\phi_{-k}\right)$$

$$\hat{q}_k \equiv \langle\hat{\boldsymbol{\Phi}}_k|\eta|\mathbf{W}_k\rangle = \int \mathrm{d}x\,\left(u_k^*(x)\phi_k^\dagger + u_k(-x)\phi_{-k}\right)$$

$$\hat{p}_k^\dagger \equiv \langle\mathbf{V}_k|\eta|\hat{\boldsymbol{\Phi}}_k\rangle = \int \mathrm{d}x\,\left(u_k^*(x)\phi_k + u_k(-x)\phi_{-k}^\dagger\right) = \hat{p}_{-k}$$

$$\hat{q}_k^\dagger \equiv \langle\mathbf{W}_k|\eta|\hat{\boldsymbol{\Phi}}_k\rangle = \int \mathrm{d}x\,\left(u_k(x)\phi_k + u_k^*(-x)\phi_{-k}^\dagger\right) = \hat{q}_{-k}$$

with $[\hat{q}_k, \hat{p}_{-k}] = i$ and other commutators zero and $\hat{a}_k = (\hat{q}_k + i\hat{p}_k)/\sqrt{2}$, accordingly $\hat{a}_k^\dagger = \left(\hat{q}_k^\dagger - i\hat{p}_k^\dagger\right)/\sqrt{2} = (\hat{q}_{-k} - i\hat{p}_{-k})/\sqrt{2}$ and $\left[\hat{a}_k, \hat{a}_k^\dagger\right] = 1$ with other commutators zero. Other choices of the phase between $\mathbf{V}_k$ and $\gamma\mathbf{V}_{-k}^*$ yield equivalent forms of the squeezing Hamiltonian [14] such as $\Gamma_k\left(\hat{a}_k\hat{a}_{-k} + \hat{a}_k^\dagger\hat{a}_{-k}^\dagger\right)$. The time-dependence of the operators is then:

$$\hat{a}_k(t) = \cosh(\Gamma_k t)\hat{a}_k(0) - i\sinh(\Gamma_k t)\hat{a}_{-k}^\dagger(0),$$
$$\hat{a}_{-k}^\dagger(t) = \cosh(\Gamma_k t)\hat{a}_{-k}^\dagger(0) + i\sinh(\Gamma_k t)\hat{a}_k(0). \tag{6}$$

**Structure factor**

The structure factor $S_k$ is obtained as follows. From the density operator $\hat{n}(x,y)$ we obtain the density fluctuation operator, only retaining the contribution from unstable modes

$$\delta\hat{n}(x,y) = \psi_0(x)\frac{1}{\sqrt{L}}\sum_k\left(\bar{u}_k(x)e^{iky}\hat{a}_k + \bar{v}_k^*(x)e^{iky}\hat{a}_{-k}^\dagger + \bar{u}_k^*(x)e^{-iky}\hat{a}_k^\dagger + \bar{v}_k(x)e^{-iky}\hat{a}_{-k}\right). \tag{7}$$

where $\bar{u}_k = (u_k^* - iu_k)/\sqrt{2}$ and $\bar{v}_k = -(u_k - iu_k^*)/\sqrt{2}$. Integrating along $x$ and taking the Fourier transform along $y$ yields the Fourier component of the one-dimensional density profile with a wavevector $q$ in the $y$-direction,

$$\delta\hat{n}_q = \int \mathrm{d}x \int \mathrm{d}y\, \delta\hat{n}(x,y)e^{-iqy}$$
$$= \sqrt{n_{1\mathrm{D}}}\int \mathrm{d}x\, \tilde{\psi}_0(x)\left[(\bar{u}_q(x) + \bar{v}_{-q}(x))\hat{a}_q + (\bar{u}_{-q}^*(x) + \bar{v}_q^*(x))\hat{a}_{-q}^\dagger\right] \tag{8}$$

Here we define $\psi_0 = \sqrt{n_{1\mathrm{D}}}\tilde{\psi}_0$ such that $\int \mathrm{d}x\,|\tilde{\psi}_0(x)|^2 = 1$. We also have $\int \mathrm{d}x\,(|\bar{u}_k(x)|^2 - |\bar{v}_{-k}(x)|^2) = 1$. The structure factor is defined as [15]

$$S_q = \frac{1}{N}\langle\delta\hat{n}_q\delta\hat{n}_q^\dagger\rangle$$
$$= \langle\left(A_q\hat{a}_q + A_q^*\hat{a}_{-q}^\dagger\right)\left(A_q^*\hat{a}_q^\dagger + A_q\hat{a}_{-q}\right)\rangle$$
$$= |A_q|^2\left(\langle 1 + \hat{a}_q^\dagger\hat{a}_q + \hat{a}_{-q}^\dagger\hat{a}_{-q}\rangle + \frac{1}{i}\langle\hat{a}_q\hat{a}_{-q} - \hat{a}_q^\dagger\hat{a}_{-q}^\dagger\rangle\right)$$
$$= |A_q|^2\left(1 + \nu\right)\cosh(2\Gamma_k t) \tag{9}$$

where $N = L n_{1D}$ is the total atom number. Since $\langle \hat{a}_q \hat{a}_{-q} - \hat{a}_q^\dagger \hat{a}_{-q}^\dagger \rangle$ is an expectation value of an operator which commutes with the Hamiltonian, it is a constant of motion and here taken to be zero. The terms $\langle \hat{a}_q^\dagger \hat{a}_q \rangle$ and $\langle \hat{a}_{-q}^\dagger \hat{a}_{-q} \rangle$ correspond to the occupation numbers of modes $\pm q$. They are related to their values at $t = 0$ using the operator time-dependence given in Eq. (6), and we denote the initial mode populations by $\nu = \langle \hat{a}_q^\dagger(0) \hat{a}_q(0) \rangle + \langle \hat{a}_{-q}^\dagger(0) \hat{a}_{-q}(0) \rangle$. The contribution of a single quantum to $S_k$ is determined by the overlap integral $A_q = A_{-q} = \int dx \, \tilde{\psi}_0(x)(\bar{u}_q(x) + \bar{v}_{-q}(x))$.

### Limit of the Lowest Landau Level

The Bogoliubov analysis in the lowest Landau level was performed in [12], focussing on the stable regime occurring in a rotating saddle potential $V(x,y) = \frac{1}{2} m \epsilon \omega^2 (x^2 - y^2)$ (in the rotating frame coordinates). The rotation frequency was chosen such that the centrifugal force precisely cancelled the trapping force in the weaker ($y$-)direction, i.e. $\Omega = \omega\sqrt{1-\epsilon}$. The experiment performed here corresponds to no rotating saddle at all, i.e. $\epsilon = 0$. The Bogoliubov Hamiltonian is

$$\hat{K}_2 = \sum_k \left( \frac{\hbar^2 k^2}{2m^*} + 2g^* n_{1D} e^{-k^2 l_B^2/2} \right) \hat{a}_k^\dagger \hat{a}_k + \frac{g^* n_{1D}}{2} \sum_k e^{-k^2 l_B^2} \left( \hat{a}_k^\dagger \hat{a}_{-k}^\dagger + \hat{a}_k \hat{a}_{-k} \right)$$

with an effective 1D coupling constant $g^* = g/\sqrt{2\pi} l_B$ and where the effective mass $m^*$ of excitations is given by $1/m^* = 1/m \left( 1 - \frac{4\Omega^2}{\omega_c^2} \right)$ with the cyclotron frequency $\omega_c = \omega\sqrt{4-2\epsilon}$ modified by the anharmonic potential. One has $m^* \approx \frac{2}{\epsilon} m$ for small $\epsilon$. We see that in the case relevant to the present experiment $\epsilon = 0$ we have $1/m^* = 0$, corresponding to "infinitely heavy" excitations, i.e. a flat band without a kinetic mass term. Importantly, although only contact interactions are present, evolution in the rotating frame yields a $k$-dependent effective interaction, and correspondingly a magneto-roton minimum which evolves into a dynamical instability as the anharmonicity $\epsilon$ decreases. The excitation spectrum follows as [12]

$$\epsilon_k^2 = \left[ \frac{\hbar^2 k^2}{2m^*} + g^* n_{1D} \left( 2 e^{-k^2 l_B^2/2} - 1 \right) \right]^2 - g^{*2} n_{1D}^2 e^{-2k^2 l_B^2} \qquad (10)$$

In the limit $\epsilon = 0$ one has an unstable Goldstone branch between $k = 0$ and $k = k_c = \sqrt{2 \log\left(\frac{1}{\sqrt{2}-1}\right)}/l_B = 1.33/l_B$. The maximum growth rate of the instability occurs at $k_{\max} l_B = \sqrt{2 \log\left(\frac{2}{\sqrt{5}-1}\right)} = 0.98$ and is $\Gamma_{k_{\max}} = \sqrt{\frac{5}{2}\sqrt{5} - \frac{11}{2}} \, g^* n_{1D} = 0.3 \, g^* n_{1D} = 0.21 g n$. This is shown in Fig. 3a).

### Evolution from stable magneto-roton excitations to dynamical instability

The expression Eq. 10 allows us to follow the excitation spectrum in the 1D regime of motion in a rotating anharmonic saddle as $\epsilon \to 0$. This evolution is shown in Fig. 5, varying the parameter introduced in [12] $\beta = \frac{n_{1D} g^*}{\hbar^2/2m^* l_B^2}$, comparing the interaction energy to the kinetic energy of excitations at momentum $\sim 1/\ell_B$. The present experiment corresponds to $\beta = \infty$, i.e. zero kinetic energy, infinite effective mass of excitations and purely interaction-driven dynamics. The figure shows how an initially stable branch consisting of phonons at low momenta $k$ develops a magneto-roton minimum at $k \approx 1/l_B$. This minimum lowers in energy as it becomes more and more favorable to create magneto-rotons, excitations which avoid the condensate mean-field repulsion due to their spatial shift by $k l_B^2 \approx l_B$. Beyond a critical $\beta = 4.9$, a dynamical instability near $k \sim 1/l_B$ develops, corresponding to the onset of magneto-roton condensation - in analogy to roton condensation considered in [16, 17]. Eventually, for $\beta \to \infty$, the case of the present experiment, the entire Goldstone branch up to $k = k_c$ is dynamically unstable, with maximum growth at $k = k_{\max} \sim 1/l_B$.

### Thomas-Fermi limit - Hydrodynamics

The Gross-Pitaevskii equation for the wavefunction $\psi = \sqrt{\rho} e^{iS}$ can be equivalently rewritten as hydrodynamic equations for the density $\rho = |\psi|^2$ and the velocity $\mathbf{v} = \frac{\hbar}{m} \nabla S$. The equation for the velocity is:

$$\frac{\partial \mathbf{v}}{\partial t} = -\nabla \left( -\frac{\hbar^2}{2m^2 \sqrt{\rho}} \nabla^2 \sqrt{\rho} + \frac{1}{2} \mathbf{v}^2 - \mathbf{v} \cdot (\mathbf{\Omega} \times \mathbf{r}) + \frac{1}{m} U + \frac{g \rho}{m} \right). \qquad (11)$$

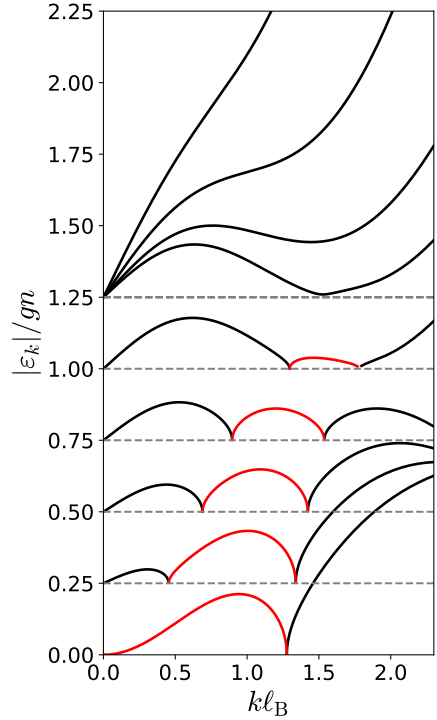

FIG. 5. Evolution from stable phonon and magneto-roton excitations to dynamically unstable excitations as the anharmonicity $\epsilon$ of a rotating saddle potential is reduced to zero. From top to bottom, the value of the parameter $\beta$, measuring the strength of interaction energy to kinetic energy, is varied from $\beta = 1, 2, 3.5, 4.9$ (the critical value where the magneto-roton minimum touches zero [12]), $5.1, 7, 10, 20, 100000$.

Here, $U(\vec{r},t) = \frac{1}{2}m\omega^2\left(x^2(1+\epsilon) + y^2(1-\epsilon)\right)$ is a rotating anisotropic potential - in the experiment $\epsilon = 0$. Introducing the "convective derivative" moving with a fluid element,

$$\frac{D}{Dt} = \frac{\partial}{\partial t} + (\mathbf{v}_{\text{rot}} \cdot \nabla) \quad (12)$$

this can be written as

$$m\frac{D\mathbf{v}_{\text{rot}}}{Dt} = 2m\mathbf{v}_{\text{rot}} \times \mathbf{\Omega} + m\Omega^2 \mathbf{r} - \nabla\left(-\frac{\hbar^2}{2m\sqrt{\rho}}\nabla^2\sqrt{\rho} + U + g\rho\right). \quad (13)$$

with $\mathbf{v}_{\text{rot}} = \mathbf{v} - \mathbf{\Omega} \times \mathbf{r}$ the velocity in the rotating frame. This is Newton's law in the rotating frame, featuring the Coriolis force $2m\mathbf{v}_{\text{rot}} \times \Omega$, the centrifugal force $m\Omega^2 \mathbf{r}$ and the force acting on a fluid particle derived from the quantum pressure $-\frac{\hbar^2}{2m\sqrt{\rho}}\nabla^2\sqrt{\rho}$, the mean-field potential $g\rho$, and the external potential $U$. The continuity equation in the rotating frame $\frac{\partial \rho}{\partial t} = -\nabla \cdot (\rho \mathbf{v}_{\text{rot}})$ can also be written using the convective derivative:

$$\frac{D\rho}{Dt} = -\rho \nabla \cdot \mathbf{v}_{\text{rot}}. \quad (14)$$

Linearizing the hydrodynamic equations - including the quantum pressure term - is equivalent to the Bogoliubov approach [4]. We now drop the quantum pressure term, considering the Thomas-Fermi limit.

We perturb the density and phase around stationary solutions $\rho_c$ and $S_c$

$$\begin{aligned} \rho &= \rho_c + \delta\rho \\ S &= S_c + \delta S \end{aligned} \quad (15)$$

and obtain the linearized hydrodynamic equations [18]

$$\begin{aligned} \frac{D\delta\rho}{Dt} &= -\frac{\hbar}{m}\nabla \cdot (\rho_c \nabla \delta S) \\ \hbar\frac{D\delta S}{Dt} &= -g\delta\rho \end{aligned} \quad (16)$$

## Hydrodynamics in the Landau gauge

The Landau gauge corresponds to setting $S_c = -\frac{m\Omega}{\hbar}xy$, and we specialize to $\Omega = \omega\sqrt{1-\epsilon}$, where the centrifugal potential exactly cancels the ($y$-)direction of weak confinement. The case in the experiment is $\epsilon = 0$. Neglecting the quantum pressure term (Thomas-Fermi limit), the stationary density profile is $g\rho_c(x) = \mu - \frac{1}{2}m\omega_c^2 x^2$, the velocity profile is $\mathbf{v}_c = \frac{\hbar}{m}\nabla S_c$ and we have $\mathbf{v}_{\rm rot} = \mathbf{v}_c - \mathbf{\Omega}\times\mathbf{r} = -2\Omega x\,\hat{\mathbf{y}}$. The coupled linearized hydrodynamic equations become

$$\frac{\partial\delta\rho}{\partial t} - 2\Omega x\partial_y\delta\rho = -\frac{\hbar}{m}\partial_x\left(\rho_c(x)\partial_x\delta S\right) - \frac{\hbar}{m}\rho_c(x)\partial_y^2\delta S$$
$$\frac{\partial\delta S}{\partial t} - 2\Omega x\partial_y\delta S = -\frac{1}{\hbar}g\delta\rho \qquad (17)$$

As there is no explicit dependence on $y$, one may choose $\delta\rho = \mathrm{Re}(e^{iky}\delta\rho_k)$ and $\delta S = \mathrm{Re}(e^{iky}\delta S_k)$. We omit the index $k$ in the following and find

$$\frac{\partial\delta\rho}{\partial t} - 2i\Omega kx\,\delta\rho = -\frac{\hbar}{m}\partial_x\left(\rho_c(x)\partial_x\delta S\right) + \frac{\hbar k^2}{m}\rho_c(x)\delta S$$
$$\frac{\partial\delta S}{\partial t} - 2i\Omega kx\,\delta S = -\frac{1}{\hbar}g\delta\rho \qquad (18)$$

Using $g\rho_c(x) = \mu - \frac{1}{2}m\omega_c^2 x^2$ and $\partial_x g\rho_c(x) = -m\omega_c^2 x$ we have

$$\frac{\partial g\delta\rho}{\partial t} - 2i\Omega kx\, g\delta\rho = \omega_c^2 x\partial_x\,\hbar\delta S - \frac{1}{m}\left(\mu - \frac{1}{2}m\omega_c^2 x^2\right)(\partial_x^2 - k^2)\,\hbar\delta S \qquad (19)$$

Measuring rates and inverse times in units of $\omega_c$ (such as $\tilde{\Omega} = \Omega/\omega_c$), energies in units of $\hbar\omega_c$ (writing $\delta\tilde{\rho} = g\delta\rho/(\hbar\omega_c)$), and lengths in units of the Thomas-Fermi radius $R_{\rm TF} = \sqrt{2\mu/(m\omega_c^2)}$, the equations become (tildes are dropped for brevity)

$$\frac{\partial\delta\rho}{\partial t} - 2i\Omega kx\,\delta\rho = \left(x\partial_x - \frac{1}{2}(1-x^2)(\partial_x^2 - k^2)\right)\delta S$$
$$\frac{\partial\delta S}{\partial t} - 2i\Omega kx\,\delta S = -\delta\rho \qquad (20)$$

The operator $\mathcal{L} \equiv x\partial_x - \frac{1}{2}(1-x^2)\partial_x^2 = -\frac{1}{2}\partial_x\left((1-x^2)\partial_x\right)$ is Legendre's differential operator, whose eigenfunctions are the Legendre polynomials $P_n(x)$:

$$\mathcal{L}P_n(x) = \frac{1}{2}n(n+1)P_n(x) \qquad (21)$$

In terms of $\mathcal{L}$, the coupled equations are

$$\frac{\partial\delta\rho}{\partial t} - 2i\Omega kx\,\delta\rho = \left(\mathcal{L} + \frac{k^2}{2}(1-x^2)\right)\delta S$$
$$\frac{\partial\delta S}{\partial t} - 2i\Omega kx\,\delta S = -\delta\rho \qquad (22)$$

Looking for a time-dependence $\sim e^{-i\omega t}$, the equations become

$$-i(\omega + 2\Omega kx)\delta\rho = \left(\mathcal{L} + \frac{k^2}{2}(1-x^2)\right)\delta S$$
$$-i(\omega + 2\Omega kx)\delta S = -\delta\rho \qquad (23)$$

and thus

$$(\omega + 2\Omega kx)^2\delta S = \left(\mathcal{L} + \frac{k^2}{2}(1-x^2)\right)\delta S \qquad (24)$$

The equation is formally solved by the confluent Heun function [19], with eigenfrequencies obtained by demanding the solution to be regular everywhere inside the Thomas-Fermi radius ($|x| < 1$). In particular, for $\epsilon = 0$ and so $2\Omega = \omega_c$ we have $\delta S = e^{\sqrt{3}kx}\,\mathrm{H}_C\left[2(k^2 + k(\sqrt{3}-2\omega) + \omega^2), 4k(\sqrt{3}-2\omega), 1, 1, 4\sqrt{3}k, \frac{1+x}{2}\right]$, where $\mathrm{H}_C[q,\alpha,\gamma,\delta,\kappa,z]$ satisfies the confluent Heun differential equation $z(z-1)y'' + (\gamma(z-1) + \delta z + z(z-1)\kappa)y' + (\alpha z - q)y = 0$ [19]. Notably, exceptional points where $\omega = 0$ are obtained as special zeroes of the confluent Heun function: $\mathrm{H}_C\left[2k^2 + 2\sqrt{3}k, 4\sqrt{3}k, 1, 1, 4\sqrt{3}k, \frac{1}{2}\right] = 0$. The critical $k = k_c$ separating the dynamically unstable modes for $0 < k < k_c$ and the stable excitations for $k > k_c$ is obtained as the first zero of this particular Heun function, at $k_c = 1.47/R_{\rm TF} = 1.04/R$. Below we find limiting cases, a series expansion for the solution, and identify the minimal set of modes responsible for the instability: The Goldstone mode, dipole mode and breathing mode of the unperturbed condensate.

*Solution for $k \to 0$*

For $k = 0$ the problem is just that of finding the excitation spectrum of a Bose-Einstein condensate in a one-dimensional harmonic oscillator:

$$\omega^2 \delta S = \mathcal{L} \delta S \tag{25}$$

with eigenvalues $\omega_n = \sqrt{\frac{1}{2} n(n+1)}$, in units of $\omega_c$, and eigenfunctions $\delta S = P_n(x)$. The case $n = 1$ is the sloshing mode, oscillating at the cyclotron frequency with $\omega_1 = \omega_c$, unshifted from the result for the non-interacting harmonic oscillator, in accordance with Kohn's theorem [10, 11]. Including the long-wavelength modulation $\propto \sin(ky)$, the mode is described by

$$\delta S(x, y, t) \propto x \cos(\omega_c t) \sin(ky)$$
$$\delta \rho(x, y, t) \propto x \sin(\omega_c t) \sin(ky)$$

which is a time-dependent "snaking" mode, sloshing back and forth along $x$ at frequency $\omega_c$.

The case $n = 2$ is the breathing mode, at $\omega_2 = \sqrt{3}\omega_c$, shifted from the non-interacting case ($2\omega_c$) by the interactions, and describing a time-dependent compression / decompression mode that periodically alternates along $y$.

The case $n = 0$ and $k = 0$ is the Goldstone mode at $\omega_0 = 0$, with $\delta S(x) = $ const. a constant phase offset, and $\delta \rho = 0$, i.e. no density modulation, describing the zero cost of changing the phase of the wavefunction globally. For the case in the experiment $\epsilon = 0$, i.e. $\Omega = \omega$, and neglecting coupling to the breathing mode, one still has the zero-energy solution $\omega_0 = 0$, since there is zero energy cost to displace the wavefunction along $x$. The mode profile is $\delta \rho(x, y, t) \propto \Omega k x \cos(ky)$, corresponding to a stationary "snake"-like deformation. With coupling to the breathing mode, it will grow exponentially.

*Solution by expansion in Legendre polynomials*

To solve the equations for $\delta \rho$ and $\delta S$, we expand them in the basis of normalized Legendre polynomials $p_n(x) \equiv \sqrt{\frac{2n+1}{2}} P_n(x)$ (with $P_n(x)$ the traditional Legendre polynomials):

$$\delta \rho(x) = \sum_n \rho_n p_n(x)$$
$$\delta S(x) = \sum_n s_n p_n(x)$$

The $p_n(x)$ are orthonormal for integration over $x \in [-1, 1]$ (while the $P_n(x)$ are not):

$$\langle n | m \rangle = \int_{-1}^{1} dx \, p_n(x) p_m(x) = \delta_{n,m}$$

where $\langle f | g \rangle = \int_{-1}^{1} dx \, f(x) g(x)$ defines a scalar product. The $p_n(x)$ are eigenfunctions of $\mathcal{L}$, but the terms in $x$ and in $x^2$ in the equations 23 couple Legendre polynomials whose index differs by 1 or 2, respectively. A recurrence relation for Legendre polynomials gives

$$x P_n = \frac{n+1}{2n+1} P_{n+1} + \frac{n}{2n+1} P_{n-1}$$

which for the orthonormal $p_n$ reads

$$x p_n = \frac{n+1}{\sqrt{(2n+1)(2n+3)}} p_{n+1} + \frac{n}{\sqrt{(2n+1)(2n-1)}} p_{n-1}$$

from which one finds

$$X_{nm} \equiv \langle n | x | m \rangle = \int_{-1}^{1} dx \, p_n(x) x p_m(x) = \frac{1}{\sqrt{(2n+1)(2m+1)}} (n \, \delta_{m,n-1} + m \, \delta_{n,m-1})$$

$$\left(X^2\right)_{nm} = \langle n|x^2|m\rangle = \sum_j \langle n|x|j\rangle \langle j|x|m\rangle = \sum_j X_{nj} X_{jm}$$

$$= \sum_j \frac{1}{\sqrt{(2n+1)(2m+1)}} \frac{1}{2j+1} (n\delta_{j,n-1} + j\delta_{n,j-1})(j\delta_{m,j-1} + m\delta_{j,m-1})$$

$$= \frac{(2n(n+1)-1)}{(2n-1)(2n+3)}\delta_{n,m} + \frac{1}{\sqrt{(2n+1)(2m+1)}}\left(\frac{m(m-1)}{(2m-1)}\delta_{m,n+2} + \frac{n(n-1)}{(2n-1)}\delta_{n,m+2}\right)$$

The equations 23 can then be written

$$\frac{\partial \rho_n}{\partial t} = i2\Omega k X_{nm}\rho_m + \mathcal{L}^{(k)}_{nm} s_m$$
$$\frac{\partial s_n}{\partial t} = i2\Omega k X_{nm} s_m - \rho_n \tag{26}$$

(using convention to sum over repeated indices) or in vector notation $\vec{\rho} = (\rho_0, \rho_1, \dots)^T$

$$\frac{\partial \vec{\rho}}{\partial t} = i2\Omega k X \vec{\rho} + \mathcal{L}^{(k)} \vec{s}$$
$$\frac{\partial \vec{s}}{\partial t} = i2\Omega k X \vec{s} - \vec{\rho} \tag{27}$$

where $X$ is the matrix with entries $X_{nm}$, and where

$$\mathcal{L}^{(k)}_{nm} = \left(\frac{1}{2}n(n+1) + \frac{k^2}{2}\right)\delta_{nm} - \frac{k^2}{2}\left(X^2\right)_{nm}$$

This linear system is solved by choosing a cutoff in the degree $n$ of polynomials used in the expansion. Alternatively, we can start with Eq. 24, which is particularly useful for the case relevant to the present experiment $\epsilon = 0$, so $2\Omega = \omega_c \equiv 1$ in dimensionless units. We have

$$(\omega + kx)^2 - \frac{k^2}{2}(1-x^2) = \omega^2 + 2\omega kx + k^2\left(\frac{3}{2}x^2 - \frac{1}{2}\right) = \omega^2 P_0 + 2\omega k P_1 + k^2 P_2$$

since $P_2(x) = \frac{3}{2}x^2 - \frac{1}{2}$. So the equation to solve is

$$\mathcal{L}\,\delta S = (k^2 P_2 + 2\omega k P_1 + \omega^2 P_0)\delta S$$

This way of writing the equation makes it explicit that the cause of the instability of the Goldstone mode ($\omega = 0$ for $k = 0$) is coupling to the breathing mode with $n = 2$, caused by $P_2$. The term multiplying $P_1$, which could in principle couple the Goldstone to the dipole mode, is zero for $\omega = 0$ and thus is not responsible for the instability. In the basis of orthonormal Legendre polynomials, we have

$$(P_2(x))_{nm} = \frac{3}{2}\left(X^2\right)_{nm} - \frac{1}{2}\delta_{nm}. \tag{28}$$

On the off-diagonals, it acts like $\frac{3}{2}(X^2)_{nm}$, but on the diagonal one finds the simpler

$$(P_2(x))_{nn} = \frac{n(n+1)}{(2n-1)(2n+3)} \tag{29}$$

This yields the equation for the $s_n$:

$$\left(\frac{1}{2}n(n+1)\left(1 - \frac{2k^2}{(2n-1)(2n+3)}\right) - \omega^2\right) s_n = 2\omega k X_{nm} s_m + \frac{3}{2}k^2 \sum_{m\neq n}\left(X^2\right)_{nm} s_m$$

$$= 2\omega k \left(\frac{n}{\sqrt{(2n-1)(2n+1)}} s_{n-1} + \frac{n+1}{\sqrt{(2n+1)(2n+3)}} s_{n+1}\right)$$

$$+ \frac{3}{2}k^2 \left(\frac{(n+1)(n+2)}{(2n+3)\sqrt{(2n+1)(2n+5)}} s_{n+2} + \frac{n(n-1)}{(2n-1)\sqrt{(2n-3)(2n+1)}} s_{n-2}\right)$$

With a finite cutoff in the polynomial degree $n$ this represents a sparse matrix, and eigenfrequencies are found by setting its determinant to zero. In general, eigenfrequencies can be complex and one finds an unstable Goldstone branch. The result is shown as the dashed lines in Fig. 3e).

*Minimal model*

Insight is obtained by truncating the Hilbert space. Including the Goldstone, dipole and breathing mode, so $p_0$, $p_1$ and $p_2$ gives equations for the coefficients $(s_0, s_1, s_2)$

$$\begin{pmatrix} \omega^2 & \frac{2k\omega}{\sqrt{3}} & \frac{k^2}{\sqrt{5}} \\ \frac{2k\omega}{\sqrt{3}} & \frac{2k^2}{5} + \omega^2 - 1 & \frac{4k\omega}{\sqrt{15}} \\ \frac{k^2}{\sqrt{5}} & \frac{4k\omega}{\sqrt{15}} & \frac{2k^2}{7} + \omega^2 - 3 \end{pmatrix} \begin{pmatrix} s_0 \\ s_1 \\ s_2 \end{pmatrix} = 0$$

This minimal model already yields a dynamically unstable Goldstone branch which will lead to exponential growth for small $k$. To find the critical $k=k_c$ where $\omega = 0$ one solves

$$\begin{vmatrix} 0 & 0 & \frac{k^2}{\sqrt{5}} \\ 0 & \frac{2k^2}{5} - 1 & 0 \\ \frac{k^2}{\sqrt{5}} & 0 & \frac{2k^2}{7} - 3 \end{vmatrix} = 0 \tag{30}$$

from which one finds $k_c R_{\text{TF}} = \sqrt{\frac{5}{2}} = 1.58$. This is already close to the exact result $k_c R_{\text{TF}} = 1.47$. The maximum instability is found at $k_{\max} R_{\text{TF}} \approx 1.18$ with $\operatorname{Im}\omega = 0.148\,\omega_c$, close to the exact maximum in the Thomas-Fermi limit at $k_{\max} R_{\text{TF}} = 1.12\ldots$ with $\operatorname{Im}\omega = 0.141\,\omega_c$.


\end{document}